\documentclass[pre,floatfix,aps,twocolumn]{revtex4-1}

\usepackage[utf8]{inputenc}
\usepackage{graphicx}
\usepackage{amsmath}
\usepackage{amsfonts}
\usepackage{bm}
\usepackage{color}
\usepackage[normal]{subfigure}
\usepackage{hyperref}

\hypersetup{colorlinks=true, citecolor=blue,
  urlcolor=black,
  pdfcreator={pdflatex},
}

\begin{document}

\title{Hyperuniformity on spherical surfaces}

\author{Ariel G. Meyra$^{1,2}$, Guillermo J. Zarragoicoechea$^{1,3}$, Alberto. L. Maltz$^4$,   Enrique Lomba$^{2}$, and Salvatore Torquato$^{5,6}$}
\affiliation{$^1$IFLYSIB (UNLP, CONICET), 59 No. 789, B1900BTE La Plata, Argentina\\
  $^2$Instituto de Qu\'{i}mica F\'{i}sica Rocasolano,
CSIC, Calle Serrano 119, E-28006 Madrid,
Spain\\
$^3$Comisión de Investigaciones Científicas de la Provincia de Buenos Aires, Argentina\\
$^4$Departamento de Matemática, Facultad de Ciencias Exactas,
Universidad Nacional de La Plata,  CC 72 Correo Central 1900 La Plata,  Argentina
\\
$^5$Department of Chemistry, Princeton University, Princeton, New Jersey 08544, USA,\\
 $^6$Princeton Institute for the Science and Technology of Materials, Princeton University, Princeton, New Jersey
 08544, USA
}
\begin{abstract}
 In this work we present a study on the characterization of ordered
 and disordered 
 hyperuniform point distributions on spherical surfaces. In spite of the extensive literature on  disordered 
 hyperuniform systems in Euclidean geometries, to date few works have dealt
 with the problem of hyperuniformity in curved spaces. As a matter of fact, some systems that
 display disordered  hyperuniformity, like the space distribution of photoreceptors
 in avian retina, actually occur  on curved surfaces. Here we
 will focus on the local particle number variance and its
 dependence on  the size of the 
sampling window (which we take
to be a spherical cap) for regular and uniform point distributions, as
well as for equilibrium configurations of fluid particles interacting
through Lennard-Jones, dipole-dipole and charge-charge potentials. We
will show how the scaling of the local number variance enables the
characterization of hyperuniform point patterns also on spherical
surfaces. 
  \end{abstract}

\maketitle

\section{Introduction}
Since the pioneering work of Torquato and Stillinger in the early 2000s
\cite{Torquato2003}, hyperuniformity has been the focus of a large
collection of works of relevance in the fields of physics (e.g. random
jammed hard-particle packings \cite{Do05d}, driven nonequilibrium
granular and colloidal systems and sand pile models
\cite{He15,We15,Tj15,Dickman2015}, and dynamical processes in ultracold
atoms \cite{Le14}), materials science (photonic band-gap
materials \cite{Fl09b,Man13b,Fr17}, dense disordered transparent
dispersions \cite{Le16}, composites with desirable transport, dielectric and fracture
properties \cite{Zh16b,Ch18,Xu17,Wu17}, polymer-grafted nanoparticle
systems \cite{Chr17}, and  ``perfect" glasses \cite{Zh17b}), and
biological systems (photoreceptor mosaics in avian retina \cite{Ji14},  and
immune system receptors \cite{Ma15}). The defining characteristic of
these hyperuniform systems is the anomalous suppression of density
(particle number or volume) variances at long wavelengths. In
Euclidean space this implies that  the structure factor $S({\bf Q})\equiv 1 + \rho {\tilde h}({\bf Q})$ tends to zero
as the wavenumber $Q \equiv |{\bf Q}| \rightarrow 0$ \cite{Torquato2003}, i.e.,
\begin{equation}
\lim_{Q\to 0} S({\bf Q}) = 0. 
\label{sq0}
\end{equation}
Here ${\tilde h}({\bf Q})$ is the Fourier transform of the total correlation
function $h({\bf r}) = g_2({\bf r})-1$, $g_2({\bf r})$ 
is the  pair correlation function and $\rho$ is the number density.

\begin{figure}[b]
\includegraphics[height=5cm,clip]{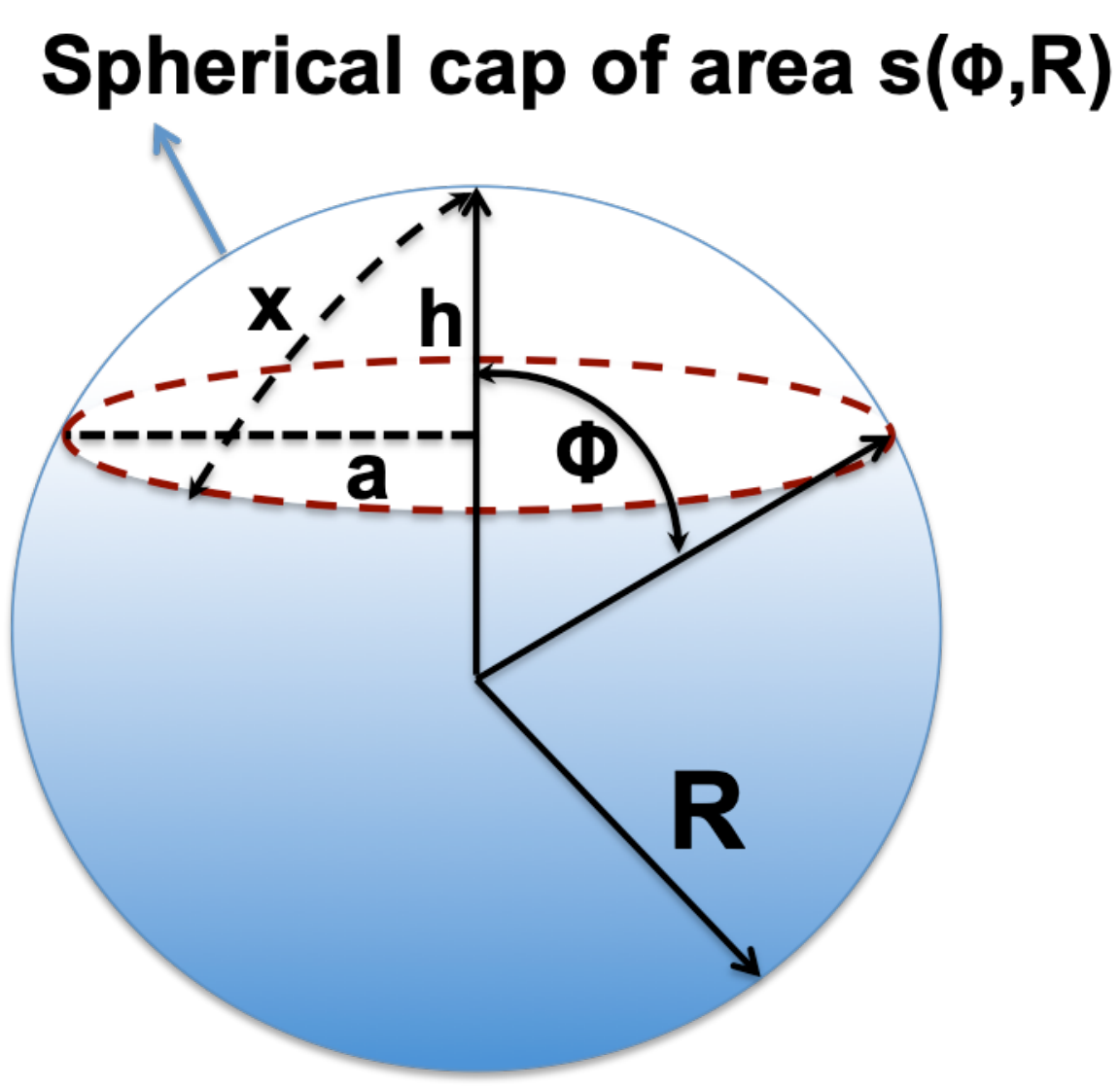}
\caption{Spherical cap sampling region (white) of arc length,$x$,
  area, $s$, and perimeter, $p$, where the number variance is calculated.}
\label{sphere}
\end{figure}

Hyperuniformity in most of the systems enumerated
above is a large scale structural property defined in an Euclidean
space. However, strictly speaking, one can also transfer the concept
to non-Euclidean geometries. A particular case of relevance in
this connection is the avian photoreceptor distribution, which, if one is
to consider it rigorously, must be treated on a curved surface, that of
the retina. Obviously, to a first approximation, if the size of the
receptors is small compared to the intrinsic curvature of the retina, one can reduce
the problem to that of a particle distribution on a flat surface.
However, this does not have to be necessarily the case in all
instances. The extension of the concept of 
 hyperuniformity to sequences of finite point sets on the sphere was
 introduced  in 
the very recent works of Brauchart and coworkers \cite{Brauchart2018,Brauchart2018a}, where the problem is addressed from a
formal mathematical perspective and connected to  the more general
problem of spherical designs. Point pattern designs on
spherical surfaces are key in the development of optimal Quasi Monte
Carlo (QMC) integration schemes \cite{Brauchart2014}. These have been
extensively used to construct efficient quadratures to evaluate
illumination integrals which are essential in the rendering of
photorealistic images \cite{Marques2015}. Brauchart and coworkers \cite{Brauchart2018} have
shown  that these optimal QMC design sequences  are hyperuniform. In
Ref.~\cite{Brauchart2014} it was shown that good candidates to build
QMC spherical designs could be devised from sets of points minimizing
Coulomb or logarithmic (i.e. two-dimensional Coulomb) pairwise
interactions. We will see here how this finding is reflected by our own
results.

On the other hand, from a materials science perspective, the
realization of particle designs on curved surfaces at the
microscopic level, has been experimentally achieved by means of
self-assembly of colloidal particles on oil/glycerol
interfaces\cite{irvine2012}. This opens an avenue to experimentally
devise and manipulate hyperuniform systems on curved surfaces at
will. Bearing in mind the relevance of hyperuniformity for the
accurate representation of images (both in bird retina and in
artificial image rendering), the potential technological implications
of these experimental achievements are obvious.

In order to further our understanding  of
hyperuniform systems in curved spaces, in this paper we have addressed the
characterization of the local particle number variances on a collection of point and
particle distributions on spherical surfaces. Given the finite size of
our systems, the use of the infinite wavelength criterion of
the structure factor  to elucidate the presence of hyperuniformity is
inadequate. This means that the structural 
 characterization of hyperuniform designs on the sphere  must be
 focused on the density/number  
 variances. Obviously,  in the limit of
infinite sphere radius with number density fixed, the
properties of the system will approximate those of the Euclidean case,
and Eq.~(\ref{sq0}) will be again useful as a signature of hyperuniformity.
This large size connection between curved and Euclidean geometries was already exploited by 
Caillol et al. \cite{Caillol1981} to remove the effects of periodic
boundary conditions in molecular simulations, and cope with the long
range of 
Coulombic interactions without resorting to the use of Ewald
summations or similar techniques.

In practice, here we will analyze the
scaling of the local particle number variance defined as
\begin{equation}
  \sigma^2_n(s) = \overline{n(s)^2} - \overline{n(s)}^2
  \label{sigma2}
\end{equation}
where $s$ denotes the area of  sampling spherical cap, and $n(s)$ is the number of particles contained in the sampling
window,  as in Ref.~\cite{Brauchart2018} (see
Figure \ref{sphere}). The bar in (\ref{sigma2}) denotes an statistical
average within the sampling window over the spherical surface. In practice, in this work we will be dealing
with point distributions composed of 
finite sets of N points placed on the surface of a sphere of radius
$R$ and total area $A=4\pi R^2$. From the work of Brauchart et
al.\cite{Brauchart2018} on hyperuniform point sets on the sphere,  we
know that for the uniform Poisson distribution  the local number
variance  scales with the surface of the sampling window, i.e.
$\sigma^2_n(s) \sim s$. In contrast, in hyperuniform systems
$\lim_{s\rightarrow\infty}\sigma^2_n(s)/s = 0$.  In Section
\ref{expl} we will introduce explicit expressions connecting the
number variance with structural properties, such as the pair
distribution function.   

 In order to properly describe
hyperuniformity on the curved sphere, in Section \ref{sample}  we 
have first analyzed the behavior of the number variance of regular point patterns on the
spherical surface, namely a triangular grid and a Fibonacci
lattice. Since spatially ordered point patterns such as those of
crystals in Euclidean space (or partly ordered, such as quasicrystals)
are known to be hyperuniform, one should clearly expect the same to
happen on the spherical surface. Next, on the opposite end, we have checked the behavior of
Poisson patterns and uniform distributions. It is important to remark
that Poisson patterns on the sphere cannot be built with a fixed
particle number,
N. For a given average surface density $\overline{\rho} = \overline{N}
/A$ one generates a series of point realizations, $i$, with a total
number of 
points, $N_i$, whose average is  $\overline{N}$ according to a
Poisson distribution (see Appendix). This implies that
we will be studying two different types of \emph{ensembles}, one
characterized by $(N,A)$ constant (\emph{canonical-like}), and another
characterized by a $(\overline{\rho},A)$ constant and variable $N$ (\emph{grand
  canonical-like}). 

In this way we will have a
set of reference systems defining ordered hyperuniform and disordered
non-hyperuniform structures.  We have then
studied the behavior of fluid particles confined on the spherical
surface and interacting via potentials with different ranges, from
short range Lennard-Jones interactions, to dipolar like (i.e. $\sim
r^{-3}$) and 3D Coulomb (plasma-like) (i.e. $\sim 1/r$)
interactions. To that aim we have run canonical Monte Carlo
simulations for various sphere sizes and a fixed surface density. 
We will see the correspondence between the hyperuniform and the
non-hyperuniform reference systems on the spherical surface and in
Euclidean space, and then we will see how the interactions and the
size of the sphere play a role in the build up of disordered
hyperuniform states on this non-Euclidean space. The article is closed
with a brief summary and future prospects. 
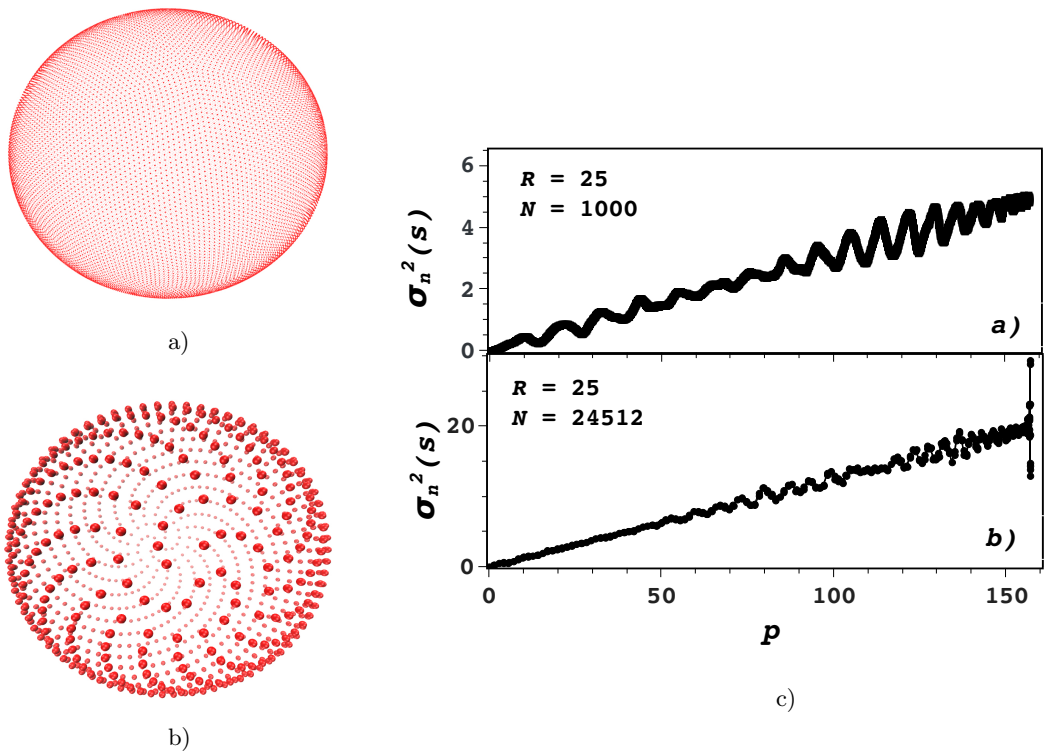
\begin{figure*}[t]
\begin{picture}(5,5)
  \put(-250,-125){\includegraphics[height=4.5cm,clip]{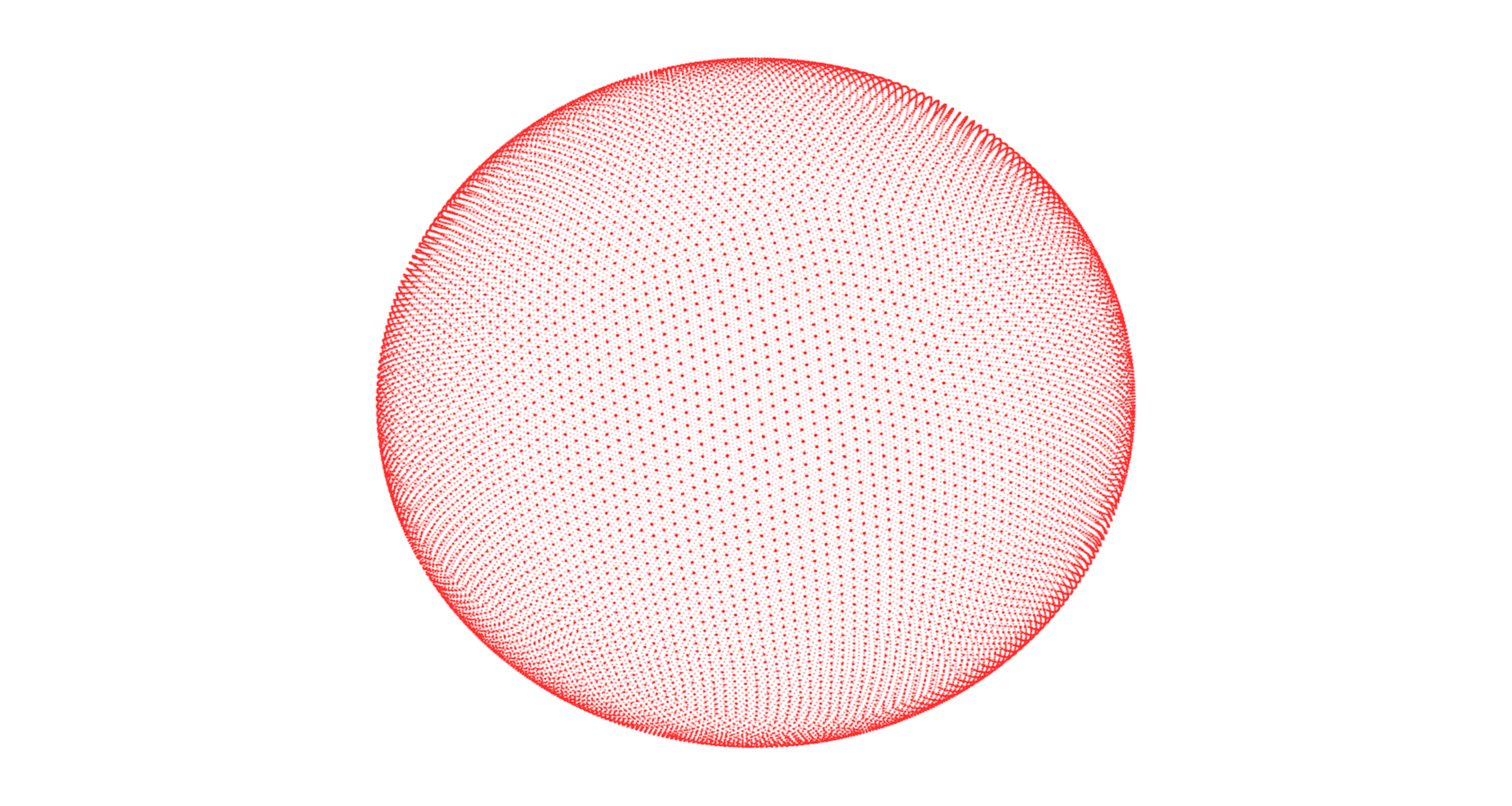}}
  \put(-130,-135){a)}

  \put(-250,-275){\includegraphics[height=4.5cm,clip]{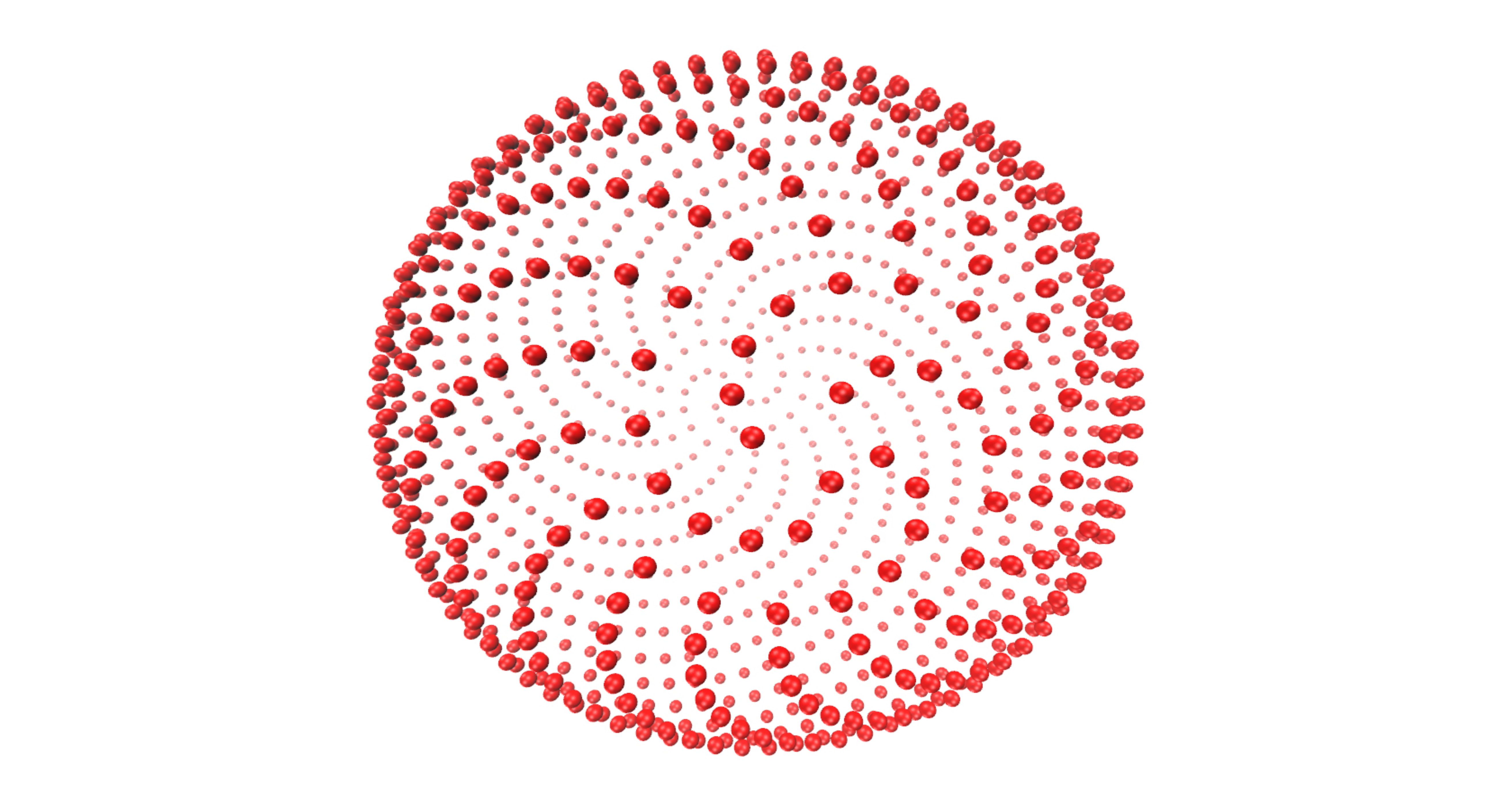}} 
  \put(-130,-285){b)}
  \put(-50,-250){ \includegraphics[width=9cm,clip]{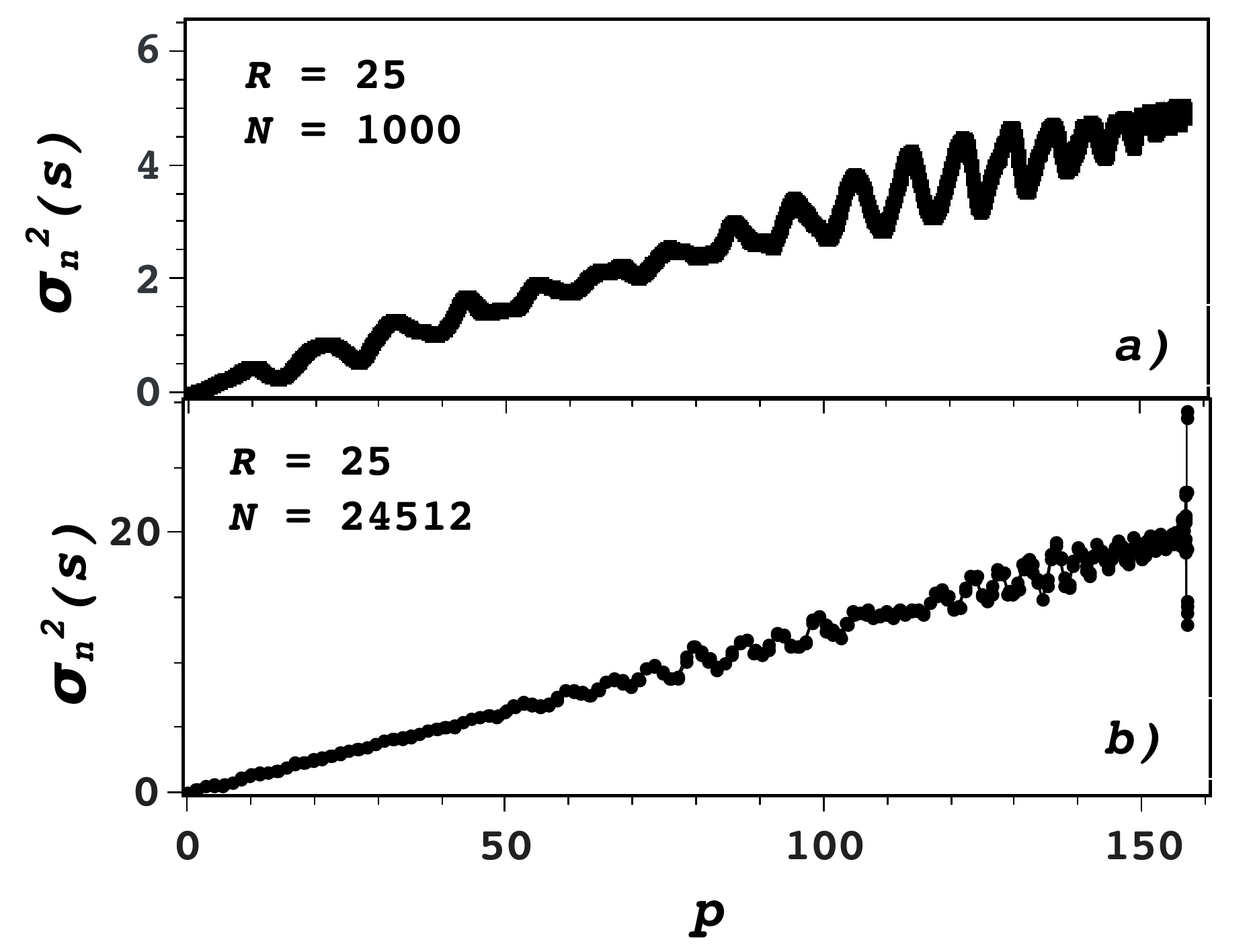}}
    \put(100,-270){c)}

\end{picture}
  \vspace{10cm}

\caption{Regular point distributions on a spherical surface: a)
  Triangular lattice on a sphere, b)Fibonacci distribution with 1000
  sample points c) Functional dependence of the
  number variance in terms of the perimeter of the sampling area,
  $p$, for the Fibonacci distribution (upper graph) and for the triangular grid (lower graph).\label{regular}} 

\end{figure*}

\section{Explicit Formulas for the Number Variance on a Sphere}
\label{expl}
Consider a {\it single} configuration of $N$ points on the 2-sphere $\mathbb{S}^2$, i.e., surface
of a three-dimensional sphere of radius $R$, as depicted in Figure \ref{sphere}. It is assumed that both $N$ and $R^2$ are large
and of comparable magnitude to one another. Let $a =\sqrt{h(2R-h)}\le
R$ and $h$ denote the base radius and the height of a spherical cap, respectively. 
The surface area of a spherical cap is $s(a) =2\pi Rh=2\pi
R^2(1-\sqrt{1-(a/R)^2})$ where we will be considering only the upper
hemisphere, $0\le\phi\le\pi/2$ to avoid ambiguity.
The number density of the points on the sphere is given by  $\rho
\equiv N/(4\pi R^2)$, and  
 $n({\bf x}_0;a)$ is the number of points contained within a spherical-cap window 
centered at position ${\bf x}_0$ on the sphere. Let the window uniformly sample
the space for sufficiently small $a$, i.e., $s(a)$ is much smaller than $2\pi R^2$.
Following Torquato and Stillinger \cite{Torquato2003} for the formulation
in Euclidean space, the number variance associated with  a single configuration on the sphere is given by
\begin{eqnarray}
{\overline {n(a)^2}}-{\overline {n(a)}}^2 
&=&  \rho s(a) \bigg[1 - \rho s(a) \nonumber \\
  & &+
\frac{1}{N}\sum_{i \neq j}^N \alpha_2(x_{ij};a)\bigg], 
\label{vol-var}
\end{eqnarray}
where ${\overline X}$ denotes again an average of a random variable $X$ over $\mathbb{S}^2$ and
$\alpha_2(x;a)$ is the intersection area of two spherical caps whose centers are separated
by a geodesic distance $x$, divided by the area of a cap.  Because the variance formula (\ref{vol-var}) is valid for a single
realization, one can use it to find the particular
point pattern that minimizes the variance at a fixed value
of $a$, i.e., one can find the ground state for the ``potential energy" function
represented by the pairwise sum in (\ref{vol-var}).

Now imagine that we generate many realizations of a large particle number $N$ on the
surface of the sphere so that the density is fixed and then consider the 
thermodynamic limit.  The ensemble-averaged number variance, $\sigma_n^2(a)$,
follows immediately from (\ref{vol-var}). We find
\begin{equation}
\sigma_n^2(a) =  \rho s(a) \Bigg[1 - \rho s(a) +
\rho \int_{\mathbb{S}^2} g_2(x) \alpha_2(x;a) d{\bf x}\Bigg] ,
\label{var}
\end{equation}
where $g_2(x)$ is the geodesic pair correlation. Brauchart
et al. \cite{Brauchart2014} rigorously studied the behavior
of the number variance in the 
large-$N$ limit. For any
``uncorrelated" point process, $g_2(x) \approx 1$ for $x \ll R$, we have
\begin{equation}
\sigma_n^2(a) =  \rho s(a),
\label{var-2}
\end{equation}
where we have used the identity $\int_{\mathbb{S}^2}  \alpha_2(x;a) d{\bf x}=s(a)$.
We call a point process on  $\mathbb{S}^2$ {\it hyperuniform} if, as $a$ becomes large,
\begin{equation}
\frac{\sigma_n^2(a)}{s(a)} \to 0.
\label{hyper}
\end{equation}

In our particular case, from Brauchart et al. \cite{Brauchart2018a},
the normalized intersection area is given by
\begin{eqnarray}
  \alpha_2(\psi;\phi) & = & 
    1 -
    \frac{1}{\pi\sin^2\phi/2}\Bigg(\arcsin(\frac{\sin\psi/2}{\sin\phi})\nonumber\\
    & &-\arcsin(\frac{\tan\psi/2}{\tan\phi})\cos\phi\Bigg)\;
    \mbox{if}\; \psi\leq 2\pi
    \label{alpha}
\end{eqnarray}
and zero otherwise, where $\psi = x/R$ is the angle between the vectors
pointing to the center of the two intersecting spherical caps.  It can
be shown that in the limit of $R\rightarrow\infty$, Eqs.~(\ref{var}) and
(\ref{alpha}) reduce to the expressions
found in Ref.~\cite{Torquato2003} for the Euclidean case in two dimensions.

\begin{figure*}[t]
\begin{picture}(5,5)
  \put(-250,-125){\includegraphics[height=4.5cm,clip]{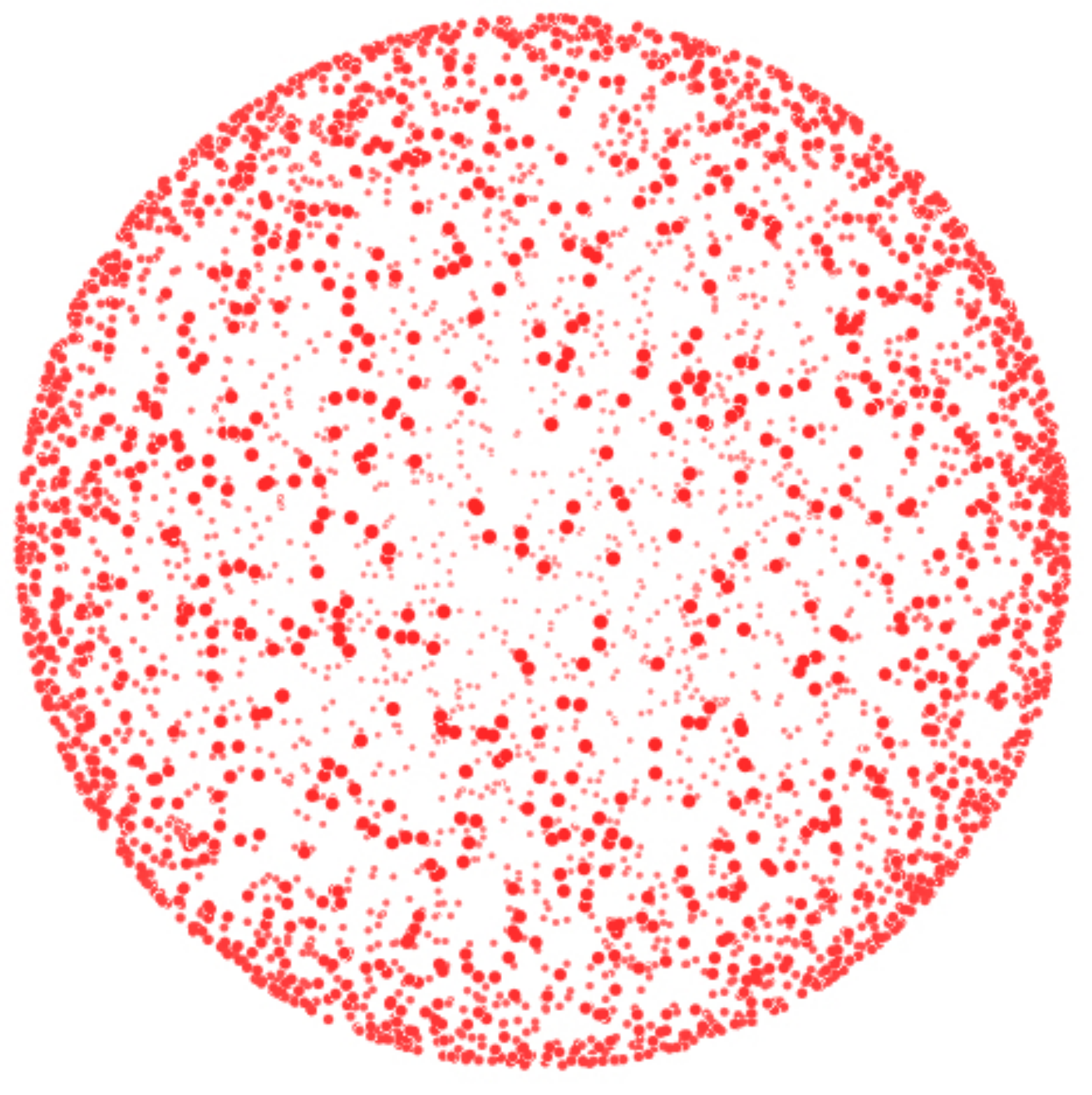}}
  \put(-130,-135){a)}

  \put(-250,-275){\includegraphics[height=4.5cm,clip]{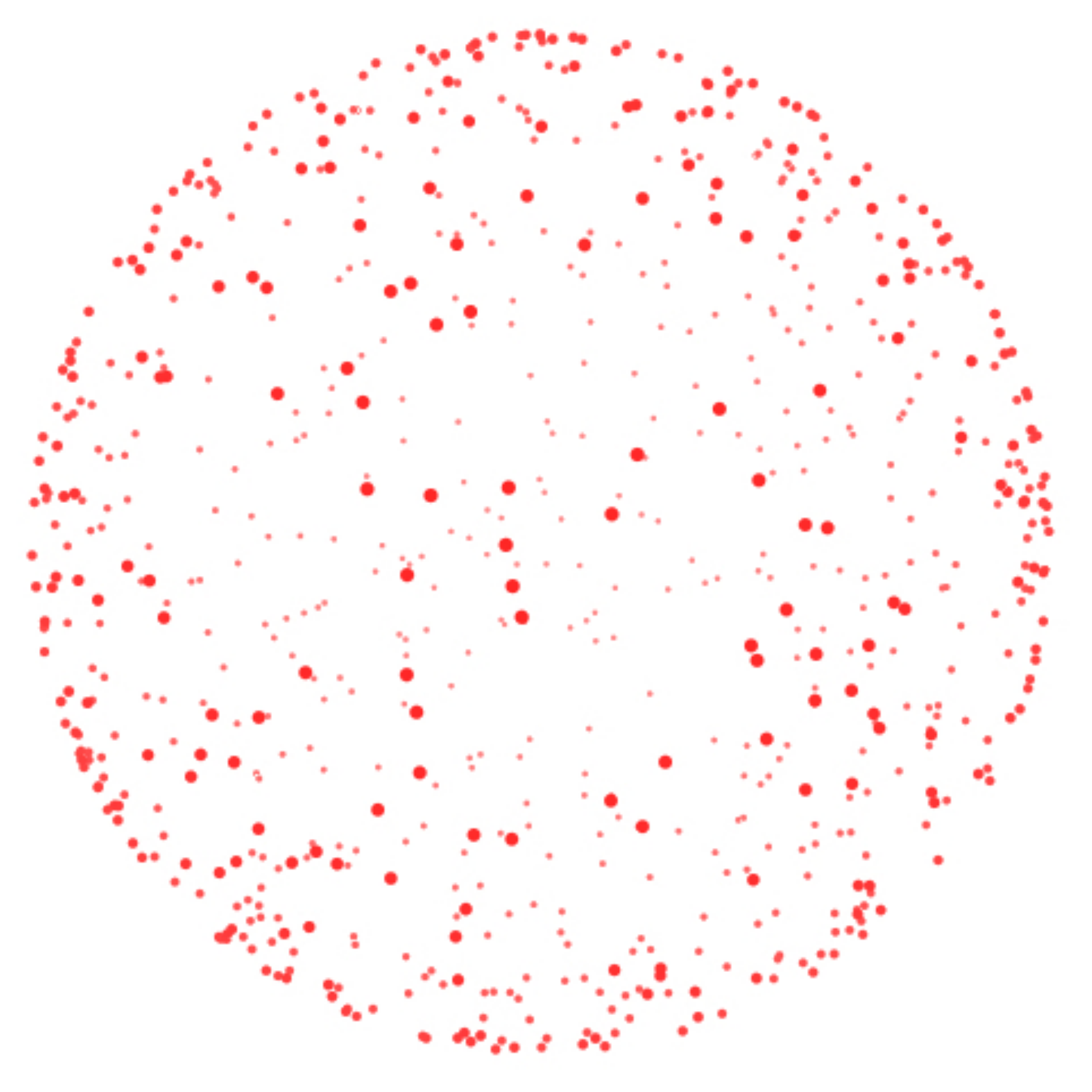}} 
  \put(-130,-285){b)}
  \put(-50,-250){ \includegraphics[width=9cm,clip]{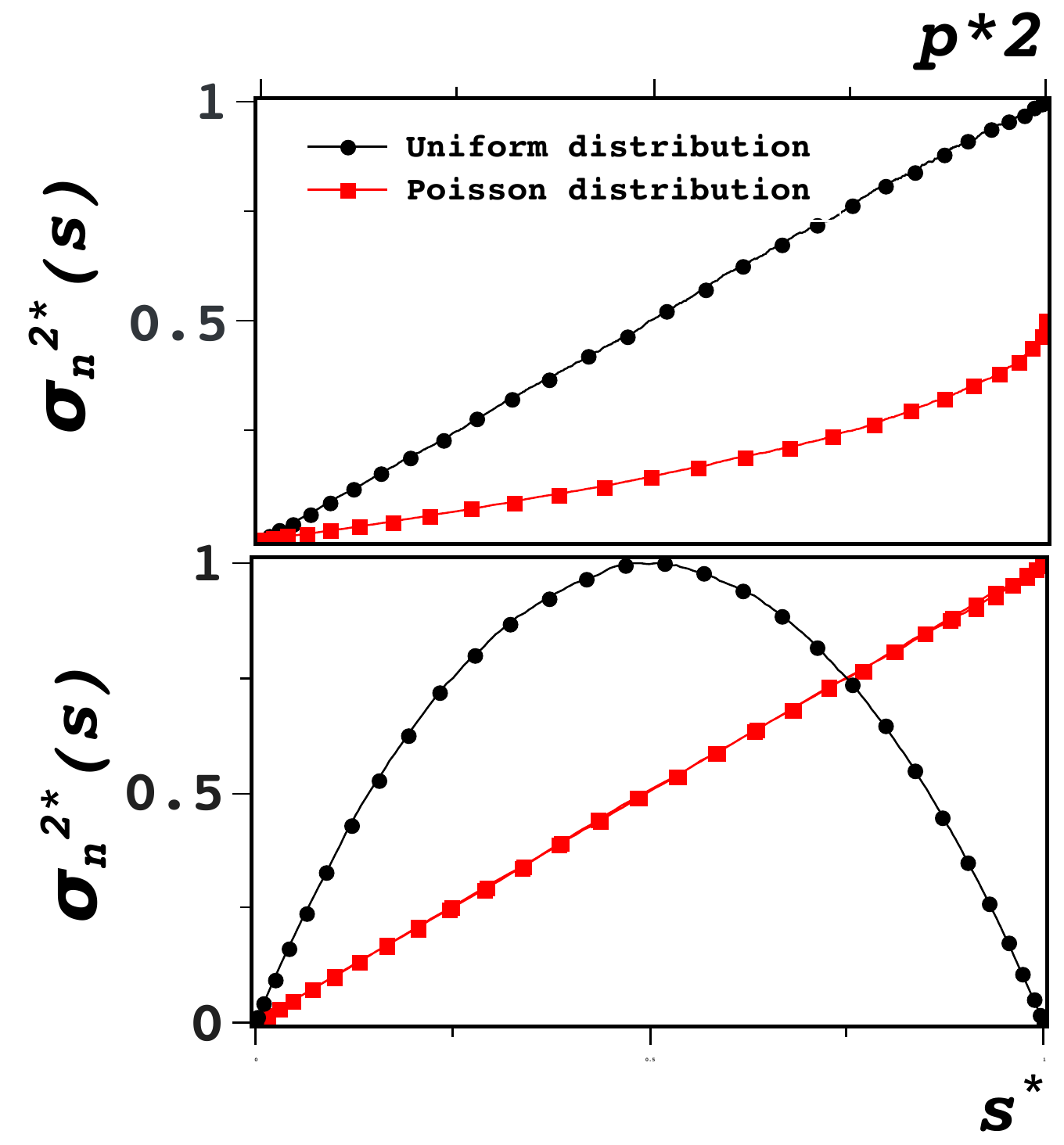}}
    \put(100,-270){c)}

\end{picture}
  \vspace{10cm}

\caption{Random point distributions on a sphere: a) uniform b)
  and Poisson (R = 1, $\lambda= 750$) c) Number variance of the
  Poisson (upper) and uniform
  distribution (lower) as functions of the normalized area, $s^*$, and
  the normalized perimeter squared, ${p^*}^2$. Note that $4s^*(1-s^*)={p^*}^2$.\label{uniform}}

\end{figure*}
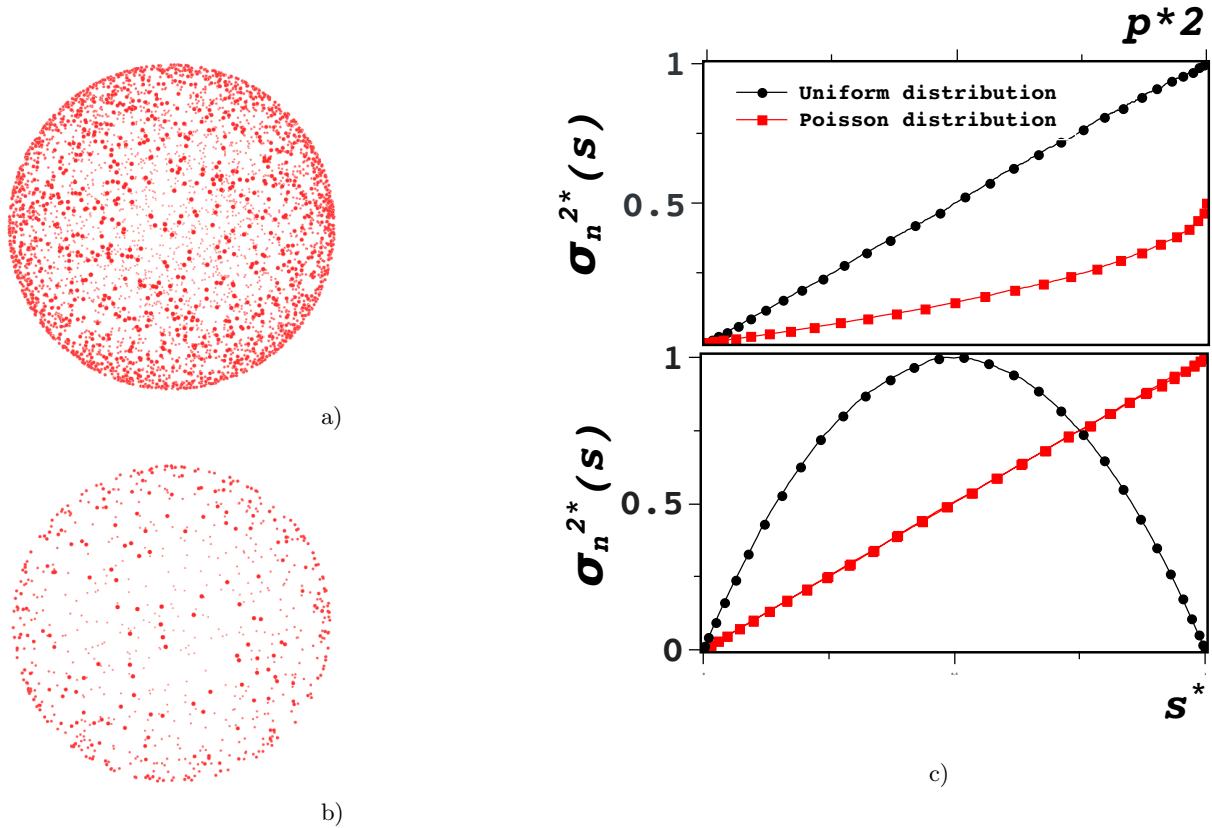

\begin{figure*}[t]
  \includegraphics[width=17cm,clip]{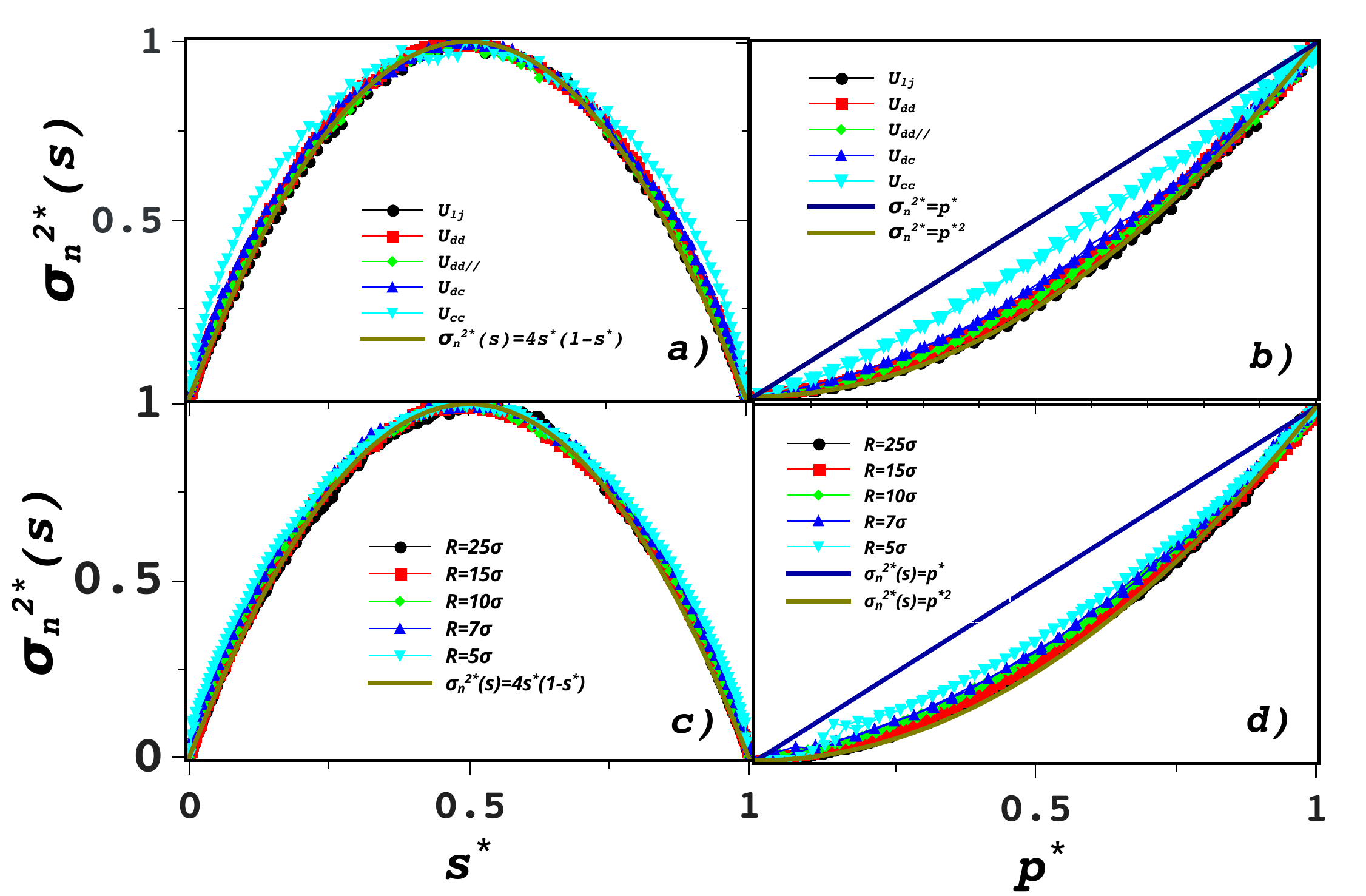}

\caption{Scaling of the normalized particle number variance,
  $\sigma_n^{*2}(s)$,  vs normalized sampling area, $s^{*}$  (a), and
  perimeter, $p^{*}$ (b),
  and density $\rho^*$= 0.5 for different interactions types and
  fixed   size ($R^*=15$) (upper
  graphs), and  for the dipolar-like interaction and various sphere
  sizes (lower graphs) . The functions $\sigma_n^{2*}(s) = 4 s^{*} (1
  - s^{*})=p^{*2}$ describes the variance dependence for a uniform
  spatial distribution of points. 
  $\sigma_n^{2*}(s) = p^{*}$ corresponds the variances of a 
  regular arrangement of points.  When the normalized
  area is used as variable, all curves apparently collapse except that
  of the 
  Coulomb interaction. Deviations from the uniform behavior are more
  apparent when the perimeter is used (right).}
\label{s2pot}
\end{figure*}

\section{Number variance of regular point patterns on a spherical surface}
\label{sample}

In this and the subsequent sections, we perform our analysis of the
local particle number variance  
defined in Eq.(\ref{sigma2}) using a spherical cap as illustrated in
Figure (\ref{sphere}). In order to perform an adequate sampling of the number variance, a
sufficiently large number of centers of the sampling spherical cap must be
chosen randomly on the surface (around 10000). In the case of fluid
particles, only three centers (in orthogonal directions) are chosen
and then averages are performed over $10^5$ configurations.
Now, since we also analyze the dependence of the results on the
total area of the system, or equivalently, the number
of particles, it will be
convenient to define normalized quantities, such as the normalized arc length
of the spherical cap, $x^{*} = x/(2\pi R)$,  the normalized area of
the sampling region,  $s^{*} = s/(4 \pi R^{2})$, and the normalized
perimeter, $p^{*}= p /(2\pi R)$. Note that these two quantities are
related by $4s^*(s^*-1) = {p^*}^2$. Also, the number variance will be
normalized as $\sigma_n^{2*}(a)$=
$\sigma_n^{2}(s)/{\sigma_n^{2}}_{\mathrm max}(s)$, where the subscript $max$
indicates the maximum value of the variance. This will facilitate
the presentation of the results and analysis of the scaling relations. 

Now, we will first consider the number variance associated with
regular point patterns (in Euclidean spaces crystals are the most
familiar regular spatial point patterns). Building a two dimensional
lattice on an spherical surface is a non trivial problem, and it is
certainly very useful in the field of astronomical observation. Here
we will resort to the icosahedron method proposed by Tegmark
\cite{Tegmark1996} as an alternative for pixelizing the celestial
sphere. We refer the reader to \cite{Tegmark1996} for the details of
the algorithm. The resulting point distribution is illustrated on
the upper left graph of Figure \ref{regular}. Note that this is an approximate
triangular grid since the algorithm maps the triangular faces of an
icosahedron in which the sphere is inscribed onto the surface of the
sphere, and then distorts the points to give all pixels approximately
the same area. Another alternative that yields equal  area
for all grid points  are the Fibonacci grids. Swinbank and
Purser have proposed an efficient algorithm to produce this very
regular grid on a spherical surface \cite{Swinbank2006}. Again, we
refer the reader to the original reference for a detailed description
of the algorithm. The corresponding illustration of the Fibonacci
pattern can be seen in the lower left graph of Figure \ref{regular}. These procedures
devised to obtain pixels of approximately the same area on the sphere
surface, are similar in spirit to the Quasi Monte Carlo approach for
numerical integration on spherical surfaces discussed by Brauchart et
al.\cite{Brauchart2018}. In the latter instance, one must choose a
set of points that minimizes the error of numerical
integration, and this in turn leads to pixels of similar size on the
sphere's surface. In Ref.~\cite{Brauchart2018} it was shown that this
corresponds  to a hyperuniform point distribution. The
minimization constraint makes the approach deterministic, retaining
nonetheless some Monte Carlo (i.e. stochastic) character. In contrast,
the result of our two tessellation techniques
would be the spherical geometry equivalent of regular grid integration
sets in Euclidean spaces.

The $\sigma_n^{2}(s)$ of the two regular point patterns is presented in
the right graph of Figure \ref{regular}. One observes in both cases the presence of
oscillations resulting from the  almost ordered structure of our
systems. On the other hand in both instances the analysis clearly
shows that we have a scaling of the
number variance linear with the perimeter, $\sigma_n^{2}(s) \propto p$. This will correspond
to the linear dependence on the radius of the sampling window in perfectly ordered lattices in a
flat two dimensional space, in which the perimeter of the sampling
window is proportional to its radius. This latter relation
does not hold on the spherical surface. In our study we have found
that no simple scaling can be
derived using the arc length, $x$, or the area, $s$, of the sampling
spherical cap. This is a necessary consequence of the finite
character of our sampling space, since when $s\rightarrow 4\pi R^2$
then $\sigma^2_n\rightarrow 0$, which implies a non monotonous
dependence on the sampling area. In contrast, $\sigma^2_n$ presents a
monotonous dependence on $p$.

\section{Variances of a random uniform and Poisson spatial
  distributions of points}
On the other end we have ``uniformly'' disordered systems, such as the
random uniform point distribution and the Poisson point distribution on
a sphere.

First, we focus on the local number variance of point patterns following a
random uniform distribution. The generation of this point
configuration  is a trivial problem in Euclidean
spaces using pseudo-random numbers. Here, one must be a bit more
careful. The simplest approach is to generate a uniform distribution
of points inside a cube inscribing the sphere, discarding those points
outside the sphere, and then performing an orthogonal projection of
the inner points  onto the surface. Alternatively one can choose three
pseudo-random numbers 
following a Gaussian distribution centered in the sphere of radius, $R$,
and project the resulting point in space onto the spherical
surface. Other approaches can also be found in
Ref. \cite{SpherePick}.  $\sigma_n^{2}(s)$   has been analyzed for
different sphere radii and  number densities and the results, once
normalized as indicated in Section \ref{sample} collapse onto a single
curve, which follows  the known exact result, $\sigma^2_n(s)^* =
4s^*(1-s^*)={p^*}^2$, as can be seen in Figure \ref{uniform}.

\begin{table}[b]

\caption{Summary of the scaling behavior of the number variance
  with the geometric 
  parameters of the sampling spherical cap whose area is $s$
  and perimeter is $p$ for regular, Poisson and uniform spatial point distributions.\label{tsum}
} 

\begin{tabular}{ll}
\hline\hline
 Point pattern & scaling\\ 
\hline
 Poisson distribution & {$\sigma_n^{2}(s)$ $\propto$ $s$} \\
 Uniform distribution & {$\sigma_n^{2}(s) \propto p^{2} \propto s(1 - s)$ }  \\
\hline
 Triangular lattice  & {$\sigma_n^{2}(s)$ $\propto$ $p$}   \\
 Fibonacci lattice  & {$\sigma_n^{2}(s)$ $\propto$ $p$}
 \\
\hline\hline
\end{tabular}

\end{table}

On the other hand, the generation of a Poisson point distribution on a spherical surface is not
straightforward.  Here it has been
generated by an algorithm devised following the considerations of Baddeley\cite{Baddeley2004}. For
completeness a detailed description of the algorithm is included in the
Appendix. A characteristic Poisson configuration on the spherical
surface is illustrated in
Figure \ref{uniform}. Note again that the configurations so generated
will be 
characterized by an average surface density, $\overline{\rho}$, and in
contrast with the previously discussed uniform distribution, we will
not have a system with a fixed number of points, $N$. Instead we will
have  a collection
of systems whose average $\overline{N}$ yields the average
density, $\overline{\rho}$. As mentioned before, to some extent, this
formulation recalls the relation between grand canonical and canonical
ensembles. In contrast with the uniform distribution, the
maximum of  the Poisson distribution variance for a given
$\overline{\rho}$ is now reached when $s\rightarrow 4\pi R^2$.

In  Figure \ref{uniform}c we present the scaling of the
normalized number variance vs the normalized area for Poisson patterns. One observes, a
complete linear dependence and the absence of oscillations
characteristic of a spatially disordered and to some extent uniform
distribution. This dependence is fully consistent with that of Poisson
patterns on flat surfaces, $\sigma_n^{2}(s) \propto s$
\cite{Torquato2003}. Interestingly, if we now try to look for the same
scaling for the the strictly uniform (constant surface density) point
distribution, one finds that it does not conform to the surface scaling
relation. In consonance with our findings for ordered, regular patterns, we
see in Figure \ref{uniform}c that
$\sigma_n^{2}(a) \propto p^2$. Thus,  it turns out that the perimeter
will now again be the scaling 
variable. In analogy with the definitions for Euclidean spaces, we
will consider systems with fixed N and  squared perimeter variance
scaling as non-hyperuniform.

In this way, we have defined what will be  our reference results for
scaling of the local number variance on the spherical
surface.  We will see
that intermediate situations between linear and quadratic scaling will
also be possible, following
\begin{equation}
  \sigma^2_n(s)  \propto p^\delta\; \mbox{with}\; 1<\delta<2,
  \label{delta}
  \end{equation}
  that from
Eq.(\ref{hyper}) will also  correspond to hyperuniform configurations. A
summary of the systems considered up to this point is presented in Table \ref{tsum}.

\section{Number variances in fluids of interacting particles}

In this section, we present some results of  Monte Carlo simulations
in a canonical ensemble (particle number, area and temperature fixed)
 for particles on a spherical surface interacting with the
potential functions summarized below in Eqs.~(\ref{lj}) and (\ref{ulr}). In what follows, $ N$,
$A$ and $T$ are the number of molecules, 
the sphere area, and temperature, respectively. The simulation starts
when N particles are randomly placed on a sphere surface of radius
$R$. We then perform $5\times 10^5$ translational attempts along 
random directions on the surface in order to equilibrate the
system. Averages are calculated over $10^5$ statistically independent
configurations.  Here we will analyze the effect of different
interactions and sphere radii (system size) on the local 
particle number variances. Bearing in mind the results of the previous
Section, we will be able to see how the interaction tunes the hyperuniform
character of the fluid structure. 

The net interparticle interaction has a short range
dispersive/repulsive component of the Lennard-Jones (LJ) form:
\begin{equation}
U_{lj}(r) = 4\epsilon \left((\frac{\sigma_{LJ}}{{r}_{ij}})^{12} - (\frac{\sigma_{LJ}}{{r}^{*}_{ij}})^{6}\right),
\label{lj}
\end{equation}
where the Lennard Jones parameters, $\epsilon$ and $\sigma_{LJ}$, are taken as the
 energy and length scale respectively. A reduced temperature is
 defined and set to $T^{*}= k T/\epsilon = 1.3$, well above the
 critical temperature for a LJ fluid practically confined in a two
 dimensional space.  Distances will be scaled as
 $r_{ij}^{*}=r_{ij}/\sigma_{LJ}$, and $R^*=R/\sigma_{LJ}$, and density
 as $\rho^*=\rho\sigma_{LJ}^2$. It must be
stressed that $r^{*}_{ij}$ is the Euclidean distance between
particles $i$ and $j$, and not the arc length. Our particles are
actually three dimensional entities with three dimensional
interactions restricted to move on the sphere's surface. To the LJ
interaction we will add dipolar-like, charge-dipole, and
charge-charge contributions. To simplify the problem, dipoles are kept
perpendicular to the surface, as if under the influence of an electric field whose source is at
the center of the sphere.
We will also consider the case of purely
parallel dipoles, which leads to a simple $1/r^3$ repulsion, and for the
charge-dipole interaction we will also consider that dipoles are
orthogonal to the line joining the particle centers. This is a crude
approximation to the case of dipoles perpendicular to the surface. The
explicit form of the interactions used is:
\begin{eqnarray}
U_{dd}& = & U_{lj} + \frac{\alpha^{*}}{(r^{*}_{ij})^{3}}[(\vec{s}_{i}\cdot\vec{s}_{j})-\frac{3(\vec{s}_{i}\cdot\vec{r}_{ij})(\vec{s}_{j}\cdot\vec{r}_{ij})}{(r^{*}_{ij})^{2}}]\nonumber\\
{U_{dd}}_{||}& = & U_{lj} + \frac{\alpha^{*}}{(r^{*}_{ij})^{3}}\nonumber\\
U_{dc} &= &  U_{lj} + \frac{\gamma^{*}}{(r^{*}_{ij})^{2}}\nonumber\\
U_{cc} &=&  U_{lj} + \frac{\beta^{*}}{(r^{*}_{ij})}\label{ulr}
\end{eqnarray}
where $\gamma^{*},\beta^{*} $ = 1 and $\alpha^*$ will be set to unity
in most cases, except when analyzing the effect of the repulsion
strength on the number variance and pattern formation.

We have determined the local number variance for two different radii,
$R^*=5$ and $R^*=15$. This is plotted for the latter case in Figure
\ref{s2pot} using normalized quantities. Results for $R^*=5$ are omitted since they are
qualitatively very similar. In the right graph of Figure \ref{s2pot} two reference curves
have been added, one representing the linear dependence on the
sampling area perimeter (strongly hyperuniform scaling) and
another for the quadratic dependence (regular disordered non-hyperuniform
systems). One immediately observes that as the range of the potential
increases, the scaling becomes hyperuniform, i.e.
  $\sigma_n^{2*}(s) \propto p^{*\delta}$ with $\delta <
2$, and $\delta$ decreasing as the interaction range increases. In
fact, for the Coulomb like interaction we have $\delta \approx 1.4$
Note the this interaction gives strictly $\delta=1$ for planar
surfaces\cite{Lomba2018}. The pure LJ fluid, as in the Euclidean case,
displays no hyperuniformity, and conforms to the same scaling as the
uniform random point patterns. 

\begin{figure}[b]
  \includegraphics[width=9cm,clip]{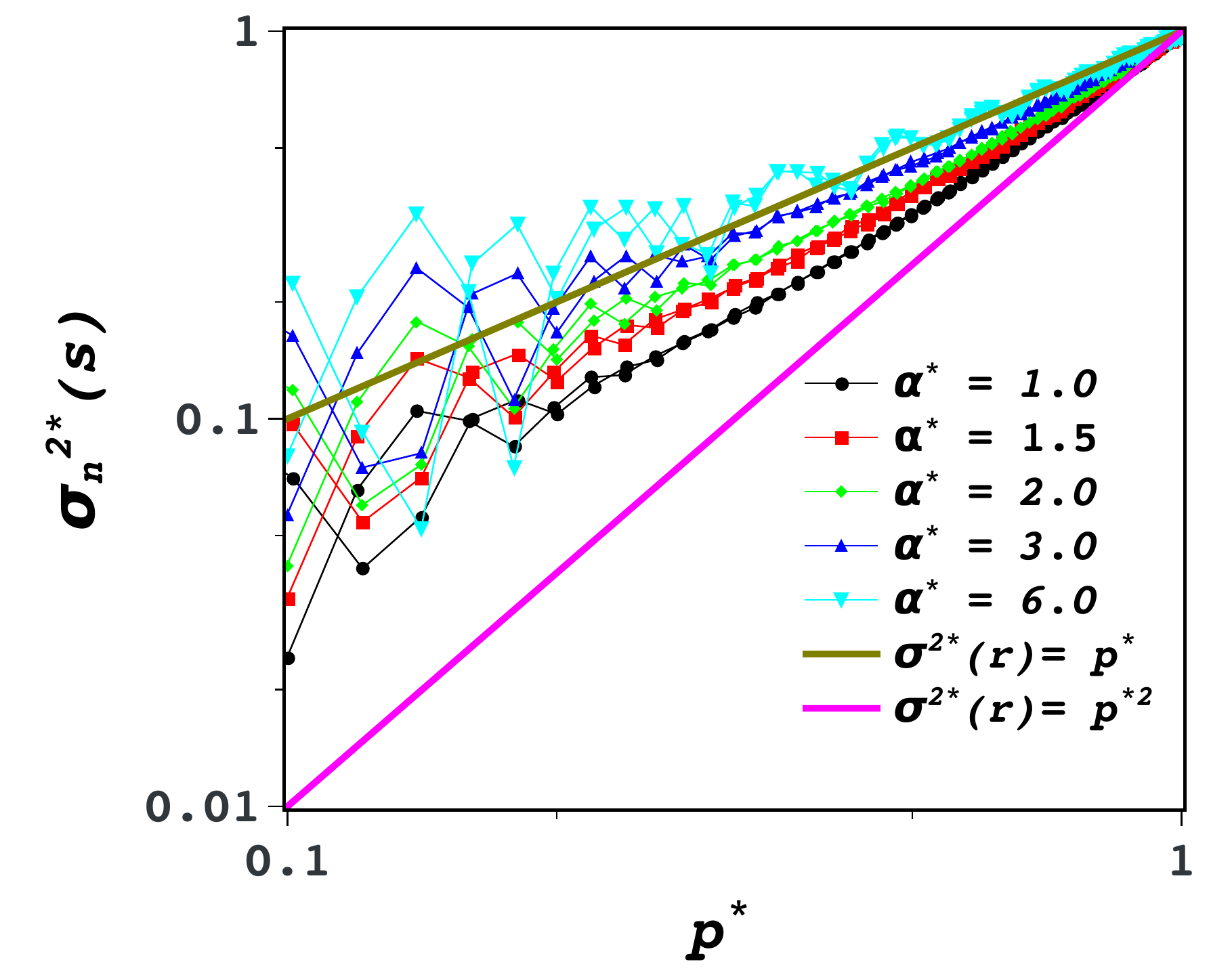}
\caption{Scaling of the normalized particle number variance,
  $\sigma_n^{*2}(s)$,  vs normalized sampling  
  perimeter, $p^{*}$  for a sphere of radius $R^* = 5$,
  and reduced density $\rho^*$= 0.5 with varying dipole-dipole
  like interparticle  repulsion.\label{strength} The double
  logarithmic scale stresses the change of regime approaching
  strongly hyperuniformity as the exponent of the curve goes to unity,
$\delta\rightarrow 1$ (see Eq.(\ref{delta})).}
\end{figure}
\begin{figure*}[t]
  \subfigure[$\alpha^{*} = 1.0$]{
    \includegraphics[width=0.25\textwidth]{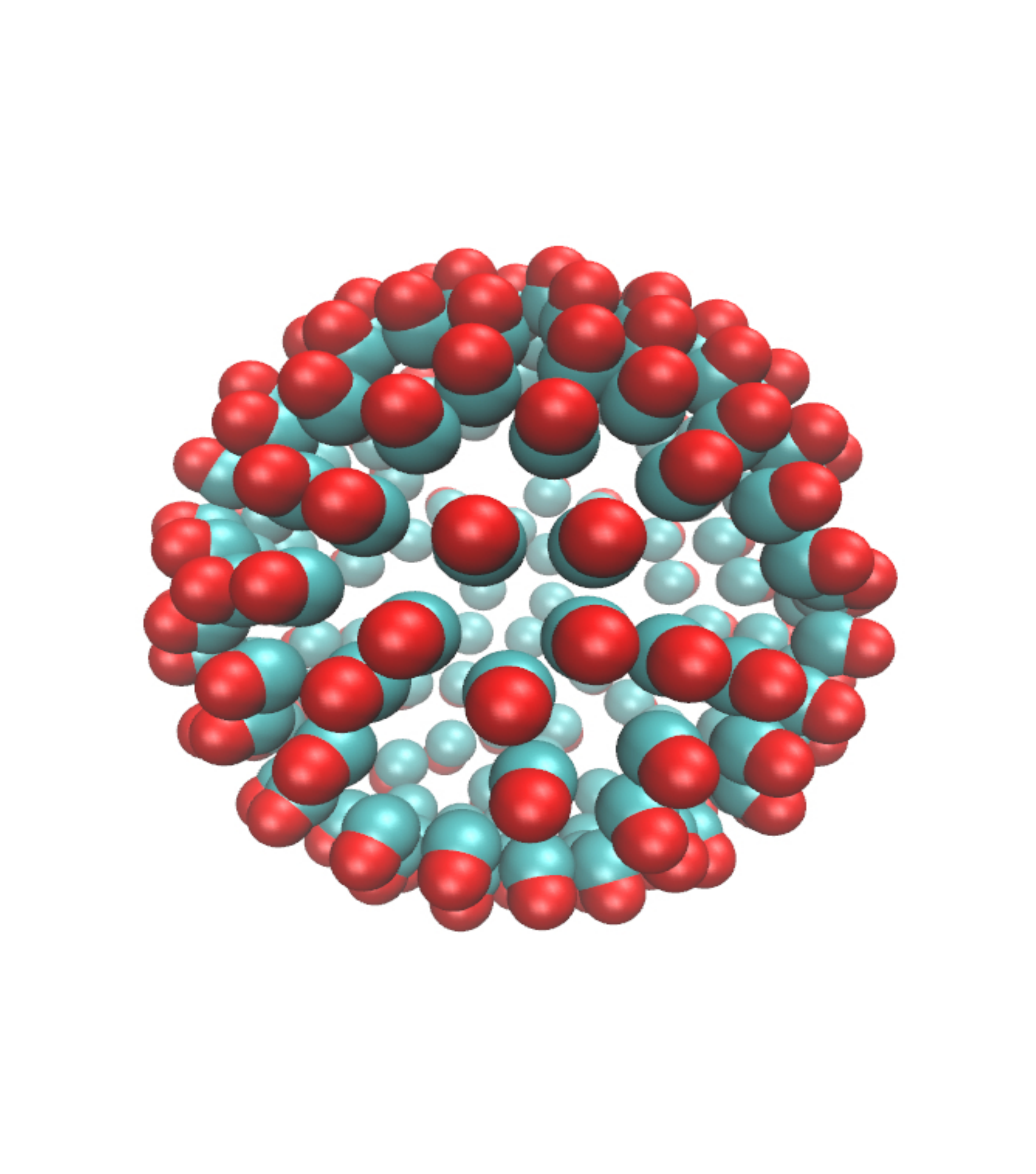} \label{f:}}
    \subfigure[$\alpha^{*} = 3.0$]{
    \includegraphics[width=0.25\textwidth]{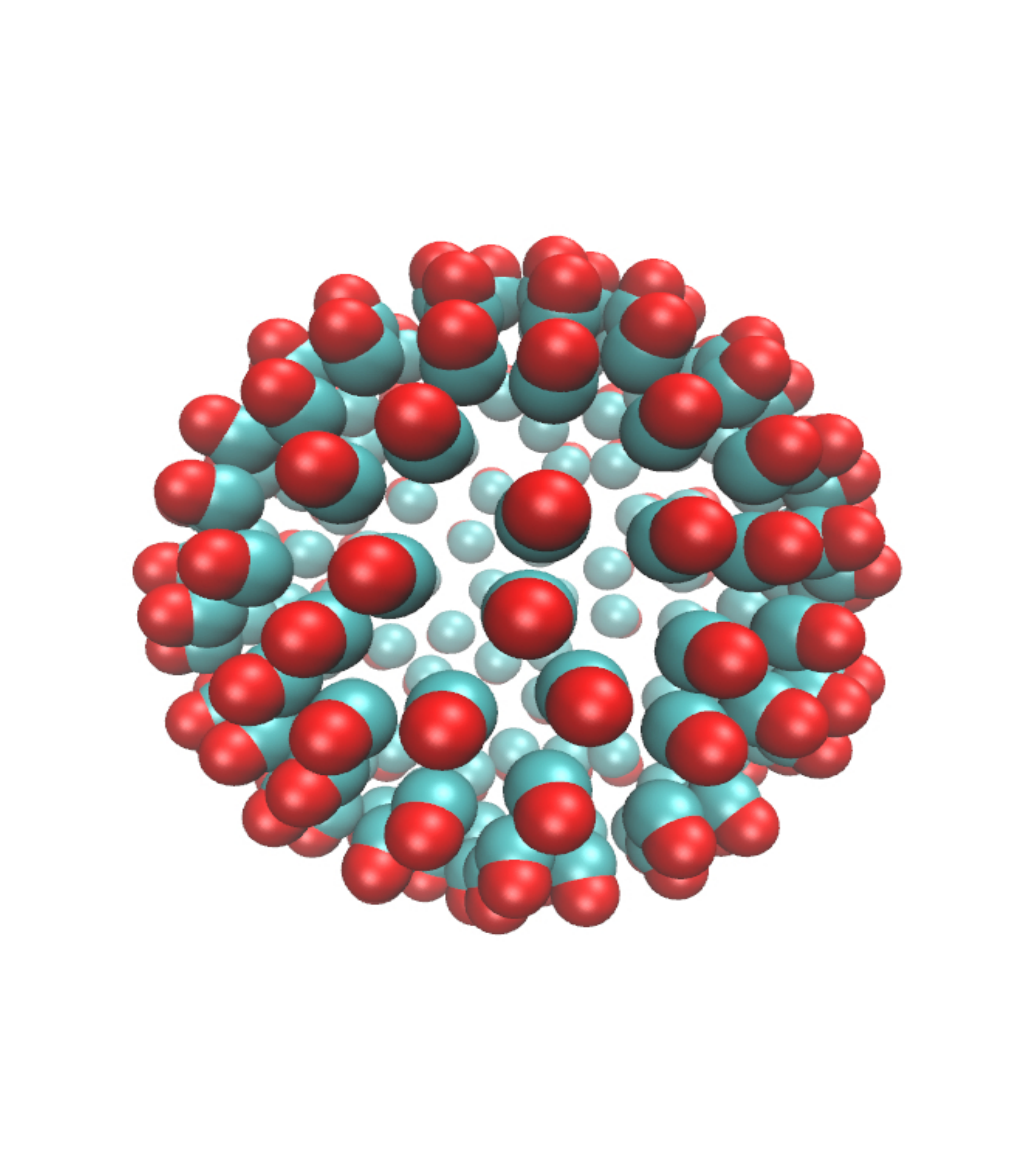}   \label{1-dipolos}}
    \subfigure[$\alpha^{*} = 6.0$]{
    \includegraphics[width=0.25\textwidth]{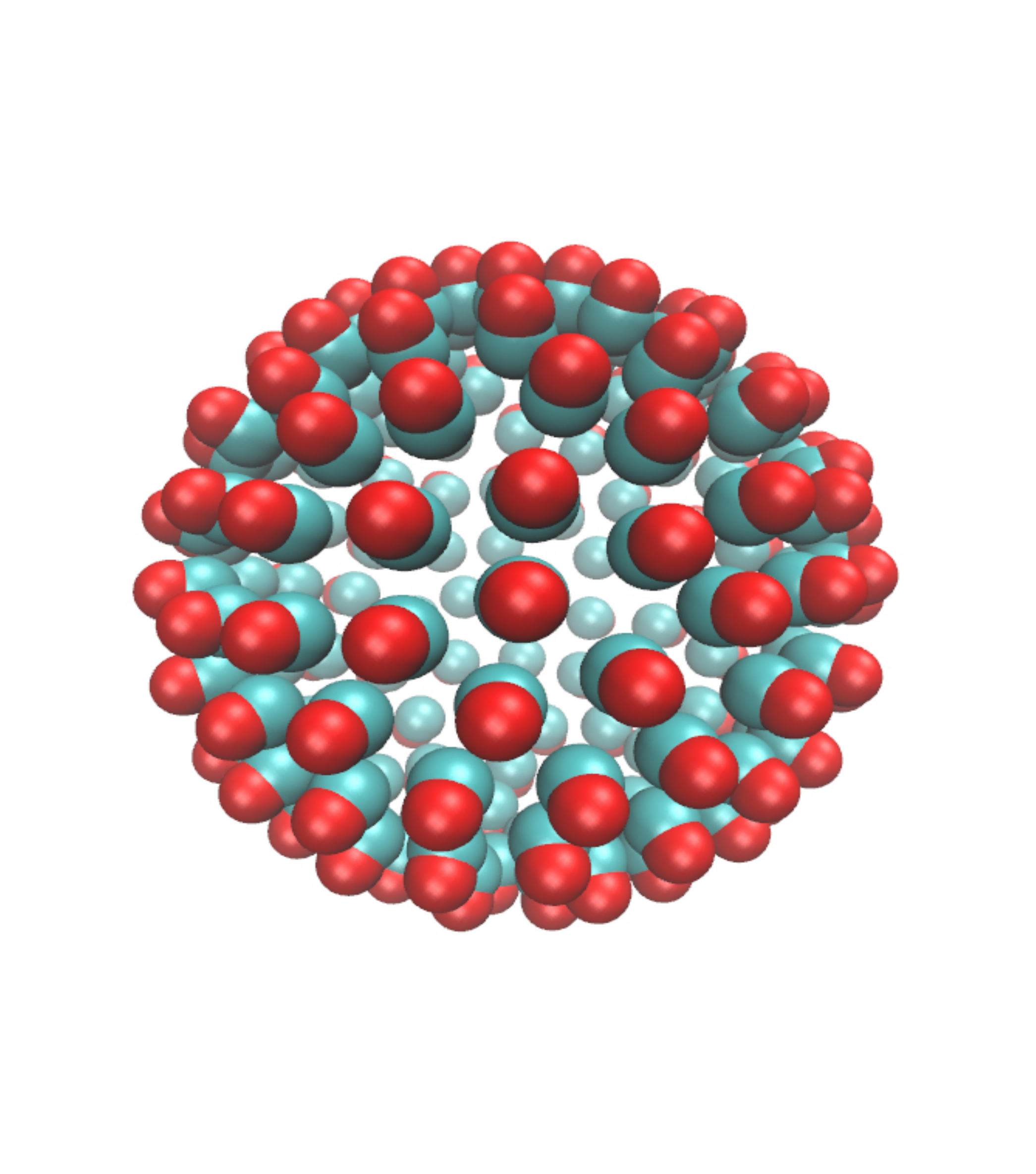}   \label{3-dipolos}}

 \caption{Snapshot of simulated system for different repulsion strength when the interaction is $U_{dd}$.}
 \label{6-dipolos}
\end{figure*}

Interestingly, a comparison of the pair distribution functions, $g_2(r)$, for
different interactions (not shown), tuning just the long range component as we
have done in the results of Figure \ref{s2pot}, does not show the
emergence of any particular feature when  hyperuniformity builds. As shown in Ref.~\cite{Lomba2017},
hyperuniformity seems to be associated with the build up of some sort
of long range order involving more than two particles, which is not
easily 
captured by pairwise quantities such as $g_2(r)$, except in the
infinite long range limit ($r\rightarrow\infty$, $Q\rightarrow 0$),
which is not accessible in the present instance.

Now, in the lower graphs of Figure \ref{s2pot} we observe how the size of the
``confining'' sphere affects the local particle number variance. To
that aim we have just focused on the dipolar-like interaction. 
The $1/r^3$ repulsion leads to a hyperuniform behavior in three
dimensions, and quasi-hyperuniformity in two dimensions (see Eq.~(4)
in Ref.~\cite{Lomba2018}). We see that for large spheres ($R^*=25$), $\sigma_n^{2*}(s)$
practically scales as a uniform random point distributions, and it
finally deviates somewhat for the smallest sphere. This is an
indication that curvature (whose relative effect is increased as the
size is smaller with fixed surface density) seems to enhance
hyperuniform-like behavior.

We turn now to the analysis of the interaction strength effects. To that purpose
we choose the dipole-dipole potential, $U_{dd}$ in Eq.(\ref{ulr}), and vary the
interaction strength parameter, $\alpha^*$,  from 1 to 6. The effect
on the local number variance is visible in Figure \ref{strength}. One
can clearly observe that as the strength of the interaction increases
(and consequently its effect on the long range order is enhanced) the
degree of hyperuniformity grows, until finally for $\alpha^*=6$ we are
back 
to the linear scaling (strong hyperuniformity) with the characteristic oscillations  of an
almost ordered regular pattern. This pattern formation is readily seen
in the 
snapshots of 
Figure \ref{6-dipolos}. One appreciates there that for the largest
interaction strength the particles are almost ordered in
a trigonal lattice. This is mostly an
 energetic effect (even if entropy is also maximized), by which the
 particles adopt a configuration that
maximizes the interparticle distances, thus minimizing the repulsive
energy.  With this
quasi-ordered state we 
are back to the purely linear dependence of the local number
variance of the trigonal and Fibonacci lattices. This low
temperature (or high 
$\alpha^*$) states recall the point patterns that minimize the
Coulomb energy, which according to Ref.~\cite{Brauchart2014} provide
suitable spherical designs for QMC integration. 

All other intermediate disordered situations are
also hyperuniform, but interestingly none of them (and neither does the
pure Coulomb repulsion) reaches at finite non-zero temperature the
limiting 
behavior  of ordered 
structures, $\delta=1$. This is in contrast with the situation found for plasmas in
Euclidean space \cite{Lomba2017,Lomba2018} which produce structural
hyperuniform configurations at any finite temperature.

\begin{figure}[b]
  \centering
\includegraphics[width=9cm,clip]{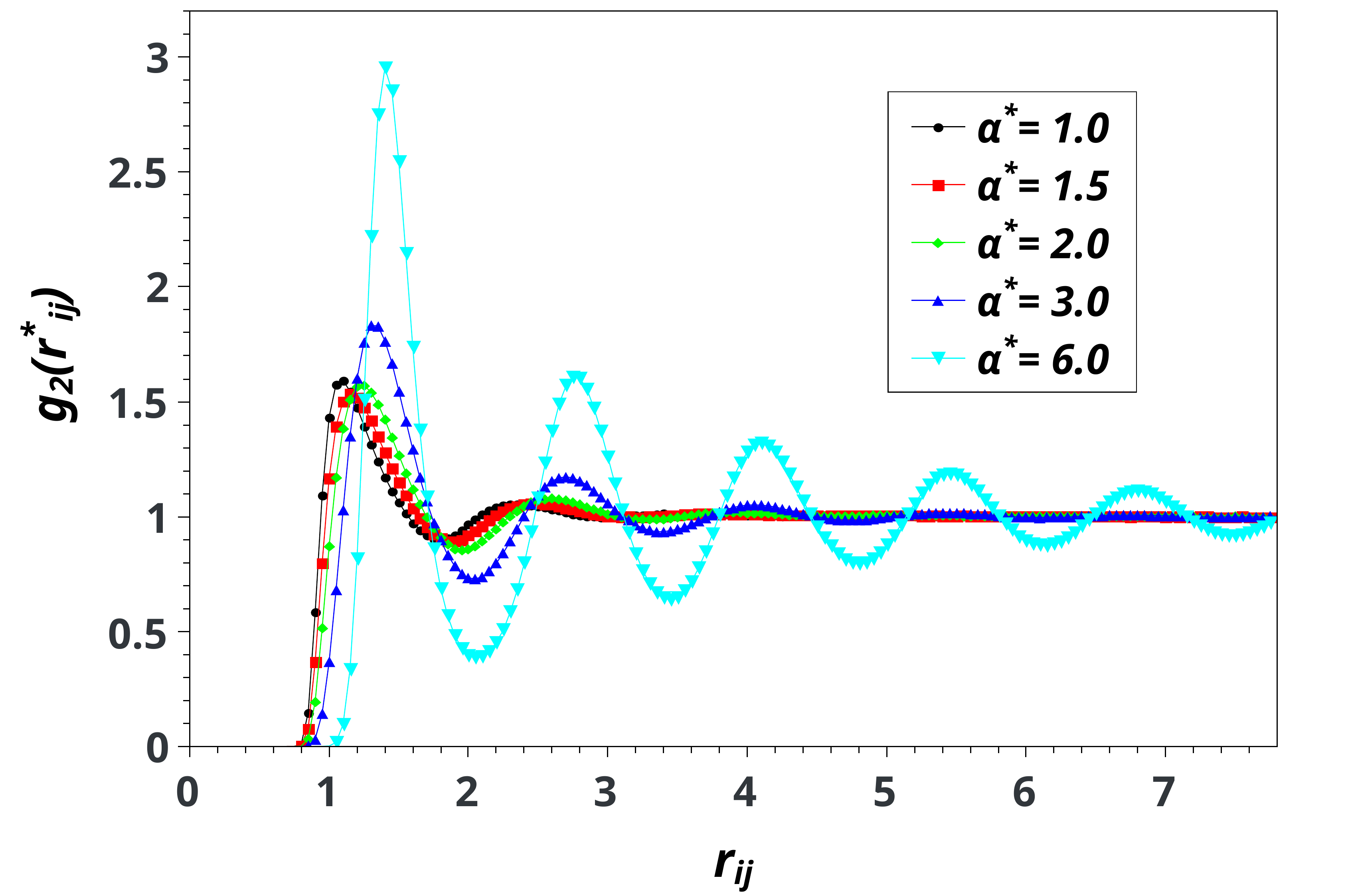}
\caption{Pair distribution function for as a function of repulsion
  strength for dipole-dipole like interactions when $R^* = 5$ and $\rho^*$= 0.5}
\label{alfa-RDF}
\end{figure}

Finally, we see now how the structuring of the fluid as a consequence of the
increasing interaction strength reflects on the pair distribution
function depicted in Figure \ref{alfa-RDF}. Here the build up of
strong short range order is seen in the marked oscillations of
$g_2(r)$ for 
$\alpha^*=6$. The fact that one obtains a smooth curve and not the
sharp spikes typical of solids is the result of the thermal
motion of the particles around the equilibrium positions and the
curved nature of the sampling space. 

\section{Conclusion}
In summary, we have shown that, in parallel with the situation found
in Euclidean space, point configurations on a sphere surface may
exhibit  two types of scaling of the local number
variance with the size of the sampling window: regular point
patterns   display a linear dependence on the perimeter of the sampling
window, and uniform point patterns show a quadratic dependence.
Strict Poisson distributions  (characterized by
an average density) have local number variances that depend
linearly on the area of the sampling 
window. Due to the curvature of the space, the area of sampling window
does not depend  quadratically on the perimeter, in contrast with
the flat two dimensional situation.
Additionally, we have seen how increasing the long range of
the interaction enhances the degree of the hyperuniformity. A similar
effect is found when decreasing the size of the sphere where the
sample is contained. This can be understood in terms of the increase
of the relative interaction range in terms of system size. Finally, we
see that for the dipolar interaction (and most likely for all
repulsive interactions of medium/long range), increasing the strength
of the interaction also induces a higher degree of hyperuniformity,
finally leading to the formation of regular ordered patterns. 

Future work will focus on the study of different geometries, such as
cylinders or  ellipsoids, of relevance in technology and biological
systems. We also plan to study the effect of interactions that favor the
formation of quasi-crystal structures, in particular those that
implement highly directional bonding interactions as found in patchy
colloids.

\bigskip

\begin{acknowledgments}
The authors are grateful to Jaeuk Kim
for his careful reading of the manuscript. AGM, GZ and EL  acknowledge the support from the European
  Union’s Horizon 2020 Research and Innovation Programme under the Marie
  Skłodowska-Curie grant agreement No 734276. EL also acknowledges
  funding  from the Agencia Estatal de
  Investigación and Fondo Europeo de Desarrollo Regional (FEDER) under
  grant No. FIS2017-89361-C3-2-P. S. T. was supported in part by the National Science Foundation under Award No. DMR-1714722.
\end{acknowledgments}

\appendix*
\section{Generating a Poisson point distribution on a sphere.}
  We recall that a random variable whose values are the non-negative integers has a
   Poisson distribution with \ parameter  $\lambda>0 $ whenever
   $P[X=k]={ e^{-\lambda} \lambda^k}/ {k!}$    for $k=1,2,...$.  It is
   often abbreviated by saying that $X$ has a  $Poiss(\lambda )$
   distribution. Some basic properties are:
    
    \begin{itemize}
    \item If $X$ has a $Poiss(\lambda )$ distribution then $E(X)=Var(X)=\lambda$
    \item If $X_1,...,X_{n}$ are independent random variables having
      $Poiss(\lambda_{1}),...,Poiss(\lambda_{n})$ distributions respectively, then
    $X_1 +... +X_{n}$  has  a $Poiss(\lambda_1 +...+\lambda_{n})$ distribution.   
    \end{itemize}          
    Let $S$ be a sphere. For each region $A\subseteq S$ we denote its area by $\mu(A)$ .
   Suppose that we have a random distribution of points on the sphere. For each region $A\subseteq S$
   we denote $N(A)$ the random variable ``number of points in $A$.''
   We have a random spatial point process with parameter   $c>0 $   whenever
   \begin{itemize}
   \item For each $A$ , $N(A)$ has a $Poiss(c\mu(A)/\mu(S))$ distribution.
   \item   If $A_1,...,A_{n}$ are mutually disjoint regions then $N(A_1),...,N(A_n)$ are
   independent random variables.                                                       
   \end{itemize}   
    We recall that each point in the sphere has two angular
    spherical coordinates $\theta$ and $\phi$.  In order to generate a set of
    points distributed according to a Poisson spatial process on the sphere with parameter $c$
    we have developed an algorithm based in an usual idea in this subject:
    \begin{enumerate}
    \item We subdivide the sphere into small, mutually disjoint
      ``spherical rectangles''   $R_1,...,R_m$  so that  
   the angular coordinates $(\theta,\phi)$ of every point in $R_j$ satisfy inequalities of
    the form $\theta_{j1}<\theta\le\theta_{j2}$ and $\phi_{j1}<\phi\le \phi_{j2}$.
    \item  For each $R_j$ we generate a random number $k_j$ according to a $Poiss(c\mu(R_j)/\mu(S))$
     distribution and   
      $k_j$ points uniformly distributed in $R_j$ are generated.  
\end{enumerate}

\bibliography{drafSPH}

\begin{thebibliography}{31}%
\makeatletter
\providecommand \@ifxundefined [1]{%
 \@ifx{#1\undefined}
}%
\providecommand \@ifnum [1]{%
 \ifnum #1\expandafter \@firstoftwo
 \else \expandafter \@secondoftwo
 \fi
}%
\providecommand \@ifx [1]{%
 \ifx #1\expandafter \@firstoftwo
 \else \expandafter \@secondoftwo
 \fi
}%
\providecommand \natexlab [1]{#1}%
\providecommand \enquote  [1]{``#1''}%
\providecommand \bibnamefont  [1]{#1}%
\providecommand \bibfnamefont [1]{#1}%
\providecommand \citenamefont [1]{#1}%
\providecommand \href@noop [0]{\@secondoftwo}%
\providecommand \href [0]{\begingroup \@sanitize@url \@href}%
\providecommand \@href[1]{\@@startlink{#1}\@@href}%
\providecommand \@@href[1]{\endgroup#1\@@endlink}%
\providecommand \@sanitize@url [0]{\catcode `\\12\catcode `\$12\catcode
  `\&12\catcode `\#12\catcode `\^12\catcode `\_12\catcode `\%12\relax}%
\providecommand \@@startlink[1]{}%
\providecommand \@@endlink[0]{}%
\providecommand \url  [0]{\begingroup\@sanitize@url \@url }%
\providecommand \@url [1]{\endgroup\@href {#1}{\urlprefix }}%
\providecommand \urlprefix  [0]{URL }%
\providecommand \Eprint [0]{\href }%
\providecommand \doibase [0]{http://dx.doi.org/}%
\providecommand \selectlanguage [0]{\@gobble}%
\providecommand \bibinfo  [0]{\@secondoftwo}%
\providecommand \bibfield  [0]{\@secondoftwo}%
\providecommand \translation [1]{[#1]}%
\providecommand \BibitemOpen [0]{}%
\providecommand \bibitemStop [0]{}%
\providecommand \bibitemNoStop [0]{.\EOS\space}%
\providecommand \EOS [0]{\spacefactor3000\relax}%
\providecommand \BibitemShut  [1]{\csname bibitem#1\endcsname}%
\let\auto@bib@innerbib\@empty
%</preamble>
\bibitem [{\citenamefont {Torquato}\ and\ \citenamefont
  {Stillinger}(2003)}]{Torquato2003}%
  \BibitemOpen
  \bibfield  {author} {\bibinfo {author} {\bibfnamefont {S.}~\bibnamefont
  {Torquato}}\ and\ \bibinfo {author} {\bibfnamefont {F.~H.}\ \bibnamefont
  {Stillinger}},\ }\href {\doibase 10.1103/PhysRevE.68.041113} {\bibfield
  {journal} {\bibinfo  {journal} {Phys. Rev. E}\ }\textbf {\bibinfo {volume}
  {68}},\ \bibinfo {pages} {041113} (\bibinfo {year} {2003})}\BibitemShut
  {NoStop}%
\bibitem [{\citenamefont {Donev}\ \emph {et~al.}(2005)\citenamefont {Donev},
  \citenamefont {Stillinger},\ and\ \citenamefont {Torquato}}]{Do05d}%
  \BibitemOpen
  \bibfield  {author} {\bibinfo {author} {\bibfnamefont {A.}~\bibnamefont
  {Donev}}, \bibinfo {author} {\bibfnamefont {F.~H.}\ \bibnamefont
  {Stillinger}}, \ and\ \bibinfo {author} {\bibfnamefont {S.}~\bibnamefont
  {Torquato}},\ }\href@noop {} {\bibfield  {journal} {\bibinfo  {journal}
  {Phys. Rev. Lett.}\ }\textbf {\bibinfo {volume} {95}},\ \bibinfo {pages}
  {090604} (\bibinfo {year} {2005})}\BibitemShut {NoStop}%
\bibitem [{\citenamefont {{Hexner}}\ and\ \citenamefont
  {{Levine}}(2015)}]{He15}%
  \BibitemOpen
  \bibfield  {author} {\bibinfo {author} {\bibfnamefont {D.}~\bibnamefont
  {{Hexner}}}\ and\ \bibinfo {author} {\bibfnamefont {D.}~\bibnamefont
  {{Levine}}},\ }\href@noop {} {\bibfield  {journal} {\bibinfo  {journal}
  {Phys. Rev. Lett.}\ }\textbf {\bibinfo {volume} {114}},\ \bibinfo {pages}
  {110602} (\bibinfo {year} {2015})}\BibitemShut {NoStop}%
\bibitem [{\citenamefont {Weijs}\ \emph {et~al.}(2015)\citenamefont {Weijs},
  \citenamefont {Jeanneret}, \citenamefont {Dreyfus},\ and\ \citenamefont
  {Bartolo}}]{We15}%
  \BibitemOpen
  \bibfield  {author} {\bibinfo {author} {\bibfnamefont {J.~H.}\ \bibnamefont
  {Weijs}}, \bibinfo {author} {\bibfnamefont {R.}~\bibnamefont {Jeanneret}},
  \bibinfo {author} {\bibfnamefont {R.}~\bibnamefont {Dreyfus}}, \ and\
  \bibinfo {author} {\bibfnamefont {D.}~\bibnamefont {Bartolo}},\ }\href@noop
  {} {\bibfield  {journal} {\bibinfo  {journal} {Phys. Rev. Lett.}\ }\textbf
  {\bibinfo {volume} {115}},\ \bibinfo {pages} {108301} (\bibinfo {year}
  {2015})}\BibitemShut {NoStop}%
\bibitem [{\citenamefont {Tjhung}\ and\ \citenamefont {Berthier}(2015)}]{Tj15}%
  \BibitemOpen
  \bibfield  {author} {\bibinfo {author} {\bibfnamefont {E.}~\bibnamefont
  {Tjhung}}\ and\ \bibinfo {author} {\bibfnamefont {L.}~\bibnamefont
  {Berthier}},\ }\href@noop {} {\bibfield  {journal} {\bibinfo  {journal}
  {Phys. Rev. Lett.}\ }\textbf {\bibinfo {volume} {114}},\ \bibinfo {pages}
  {148301} (\bibinfo {year} {2015})}\BibitemShut {NoStop}%
\bibitem [{\citenamefont {Dickman}\ and\ \citenamefont
  {da~Cunha}(2015)}]{Dickman2015}%
  \BibitemOpen
  \bibfield  {author} {\bibinfo {author} {\bibfnamefont {R.}~\bibnamefont
  {Dickman}}\ and\ \bibinfo {author} {\bibfnamefont {S.~D.}\ \bibnamefont
  {da~Cunha}},\ }\href {\doibase 10.1103/physreve.92.020104} {\bibfield
  {journal} {\bibinfo  {journal} {Physical Review E}\ }\textbf {\bibinfo
  {volume} {92}} (\bibinfo {year} {2015}),\
  10.1103/physreve.92.020104}\BibitemShut {NoStop}%
\bibitem [{\citenamefont {{Lesanovsky}}\ and\ \citenamefont
  {{Garrahan}}(2014)}]{Le14}%
  \BibitemOpen
  \bibfield  {author} {\bibinfo {author} {\bibfnamefont {I.}~\bibnamefont
  {{Lesanovsky}}}\ and\ \bibinfo {author} {\bibfnamefont {J.~P.}\ \bibnamefont
  {{Garrahan}}},\ }\href@noop {} {\bibfield  {journal} {\bibinfo  {journal}
  {Phys. Rev. A}\ }\textbf {\bibinfo {volume} {90}},\ \bibinfo {pages} {011603}
  (\bibinfo {year} {2014})}\BibitemShut {NoStop}%
\bibitem [{\citenamefont {Florescu}\ \emph {et~al.}(2009)\citenamefont
  {Florescu}, \citenamefont {Torquato},\ and\ \citenamefont
  {Steinhardt}}]{Fl09b}%
  \BibitemOpen
  \bibfield  {author} {\bibinfo {author} {\bibfnamefont {M.}~\bibnamefont
  {Florescu}}, \bibinfo {author} {\bibfnamefont {S.}~\bibnamefont {Torquato}},
  \ and\ \bibinfo {author} {\bibfnamefont {P.~J.}\ \bibnamefont {Steinhardt}},\
  }\href@noop {} {\bibfield  {journal} {\bibinfo  {journal} {Proc. Nat. Acad.
  Sci.}\ }\textbf {\bibinfo {volume} {106}},\ \bibinfo {pages} {20658}
  (\bibinfo {year} {2009})}\BibitemShut {NoStop}%
\bibitem [{\citenamefont {Man}\ \emph {et~al.}(2013)\citenamefont {Man},
  \citenamefont {Florescu}, \citenamefont {Williamson}, \citenamefont {He},
  \citenamefont {Hashemizad}, \citenamefont {Leung}, \citenamefont {Liner},
  \citenamefont {Torquato}, \citenamefont {Chaikin},\ and\ \citenamefont
  {Steinhardt}}]{Man13b}%
  \BibitemOpen
  \bibfield  {author} {\bibinfo {author} {\bibfnamefont {W.}~\bibnamefont
  {Man}}, \bibinfo {author} {\bibfnamefont {M.}~\bibnamefont {Florescu}},
  \bibinfo {author} {\bibfnamefont {E.~P.}\ \bibnamefont {Williamson}},
  \bibinfo {author} {\bibfnamefont {Y.}~\bibnamefont {He}}, \bibinfo {author}
  {\bibfnamefont {S.~R.}\ \bibnamefont {Hashemizad}}, \bibinfo {author}
  {\bibfnamefont {B.~Y.~C.}\ \bibnamefont {Leung}}, \bibinfo {author}
  {\bibfnamefont {D.~R.}\ \bibnamefont {Liner}}, \bibinfo {author}
  {\bibfnamefont {S.}~\bibnamefont {Torquato}}, \bibinfo {author}
  {\bibfnamefont {P.~M.}\ \bibnamefont {Chaikin}}, \ and\ \bibinfo {author}
  {\bibfnamefont {P.~J.}\ \bibnamefont {Steinhardt}},\ }\href@noop {}
  {\bibfield  {journal} {\bibinfo  {journal} {Proc. Nat. Acad. Sci.}\ }\textbf
  {\bibinfo {volume} {110}},\ \bibinfo {pages} {15886} (\bibinfo {year}
  {2013})}\BibitemShut {NoStop}%
\bibitem [{\citenamefont {{Froufe-P{\'e}rez}}\ \emph
  {et~al.}(2017)\citenamefont {{Froufe-P{\'e}rez}}, \citenamefont {{Engel}},
  \citenamefont {{Jos{\'e} S{\'a}enz}},\ and\ \citenamefont
  {{Scheffold}}}]{Fr17}%
  \BibitemOpen
  \bibfield  {author} {\bibinfo {author} {\bibfnamefont {L.~S.}\ \bibnamefont
  {{Froufe-P{\'e}rez}}}, \bibinfo {author} {\bibfnamefont {M.}~\bibnamefont
  {{Engel}}}, \bibinfo {author} {\bibfnamefont {J.}~\bibnamefont {{Jos{\'e}
  S{\'a}enz}}}, \ and\ \bibinfo {author} {\bibfnamefont {F.}~\bibnamefont
  {{Scheffold}}},\ }\href@noop {} {\bibfield  {journal} {\bibinfo  {journal}
  {Proc. Nat. Acad. Sci.}\ }\textbf {\bibinfo {volume} {114}},\ \bibinfo
  {pages} {9570–} (\bibinfo {year} {2017})}\BibitemShut {NoStop}%
\bibitem [{\citenamefont {{Leseur}}\ \emph {et~al.}(2016)\citenamefont
  {{Leseur}}, \citenamefont {{Pierrat}},\ and\ \citenamefont
  {{Carminati}}}]{Le16}%
  \BibitemOpen
  \bibfield  {author} {\bibinfo {author} {\bibfnamefont {O.}~\bibnamefont
  {{Leseur}}}, \bibinfo {author} {\bibfnamefont {R.}~\bibnamefont {{Pierrat}}},
  \ and\ \bibinfo {author} {\bibfnamefont {R.}~\bibnamefont {{Carminati}}},\
  }\href@noop {} {\bibfield  {journal} {\bibinfo  {journal} {Optica}\ }\textbf
  {\bibinfo {volume} {3}},\ \bibinfo {pages} {763} (\bibinfo {year}
  {2016})}\BibitemShut {NoStop}%
\bibitem [{\citenamefont {Zhang}\ \emph {et~al.}(2016)\citenamefont {Zhang},
  \citenamefont {Stillinger},\ and\ \citenamefont {Torquato}}]{Zh16b}%
  \BibitemOpen
  \bibfield  {author} {\bibinfo {author} {\bibfnamefont {G.}~\bibnamefont
  {Zhang}}, \bibinfo {author} {\bibfnamefont {F.~H.}\ \bibnamefont
  {Stillinger}}, \ and\ \bibinfo {author} {\bibfnamefont {S.}~\bibnamefont
  {Torquato}},\ }\href@noop {} {\bibfield  {journal} {\bibinfo  {journal} {J.
  Chem. Phys}\ }\textbf {\bibinfo {volume} {145}},\ \bibinfo {pages} {244109}
  (\bibinfo {year} {2016})}\BibitemShut {NoStop}%
\bibitem [{\citenamefont {{Chen}}\ and\ \citenamefont
  {{Torquato}}(2018)}]{Ch18}%
  \BibitemOpen
  \bibfield  {author} {\bibinfo {author} {\bibfnamefont {D.}~\bibnamefont
  {{Chen}}}\ and\ \bibinfo {author} {\bibfnamefont {S.}~\bibnamefont
  {{Torquato}}},\ }\href@noop {} {\bibfield  {journal} {\bibinfo  {journal}
  {Acta Materialia}\ }\textbf {\bibinfo {volume} {142}},\ \bibinfo {pages}
  {152} (\bibinfo {year} {2018})}\BibitemShut {NoStop}%
\bibitem [{\citenamefont {Xu}\ \emph {et~al.}(2017)\citenamefont {Xu},
  \citenamefont {Chen}, \citenamefont {Chen}, \citenamefont {Xu},\ and\
  \citenamefont {Jiao}}]{Xu17}%
  \BibitemOpen
  \bibfield  {author} {\bibinfo {author} {\bibfnamefont {Y.}~\bibnamefont
  {Xu}}, \bibinfo {author} {\bibfnamefont {S.}~\bibnamefont {Chen}}, \bibinfo
  {author} {\bibfnamefont {P.-E.}\ \bibnamefont {Chen}}, \bibinfo {author}
  {\bibfnamefont {W.}~\bibnamefont {Xu}}, \ and\ \bibinfo {author}
  {\bibfnamefont {Y.}~\bibnamefont {Jiao}},\ }\href@noop {} {\bibfield
  {journal} {\bibinfo  {journal} {Phys. Rev. E}\ }\textbf {\bibinfo {volume}
  {96}},\ \bibinfo {pages} {043301} (\bibinfo {year} {2017})}\BibitemShut
  {NoStop}%
\bibitem [{\citenamefont {Wu}\ \emph {et~al.}(2017)\citenamefont {Wu},
  \citenamefont {Sheng},\ and\ \citenamefont {Hao}}]{Wu17}%
  \BibitemOpen
  \bibfield  {author} {\bibinfo {author} {\bibfnamefont {B.-Y.}\ \bibnamefont
  {Wu}}, \bibinfo {author} {\bibfnamefont {X.-Q.}\ \bibnamefont {Sheng}}, \
  and\ \bibinfo {author} {\bibfnamefont {Y.}~\bibnamefont {Hao}},\ }\href@noop
  {} {\bibfield  {journal} {\bibinfo  {journal} {PloS one}\ }\textbf {\bibinfo
  {volume} {12}},\ \bibinfo {pages} {e0185921} (\bibinfo {year}
  {2017})}\BibitemShut {NoStop}%
\bibitem [{\citenamefont {Chremos}\ and\ \citenamefont
  {Douglas}(2017)}]{Chr17}%
  \BibitemOpen
  \bibfield  {author} {\bibinfo {author} {\bibfnamefont {A.}~\bibnamefont
  {Chremos}}\ and\ \bibinfo {author} {\bibfnamefont {J.~F.}\ \bibnamefont
  {Douglas}},\ }\href@noop {} {\bibfield  {journal} {\bibinfo  {journal}
  {Annalen der Physik}\ }\textbf {\bibinfo {volume} {529}} (\bibinfo {year}
  {2017})}\BibitemShut {NoStop}%
\bibitem [{\citenamefont {{Zhang}}\ \emph {et~al.}(2017)\citenamefont
  {{Zhang}}, \citenamefont {{Stillinger}},\ and\ \citenamefont
  {{Torquato}}}]{Zh17b}%
  \BibitemOpen
  \bibfield  {author} {\bibinfo {author} {\bibfnamefont {G.}~\bibnamefont
  {{Zhang}}}, \bibinfo {author} {\bibfnamefont {F.~H.}\ \bibnamefont
  {{Stillinger}}}, \ and\ \bibinfo {author} {\bibfnamefont {S.}~\bibnamefont
  {{Torquato}}},\ }\href@noop {} {\bibfield  {journal} {\bibinfo  {journal}
  {Phys. Rev. E}\ }\textbf {\bibinfo {volume} {96}},\ \bibinfo {pages} {042146}
  (\bibinfo {year} {2017})}\BibitemShut {NoStop}%
\bibitem [{\citenamefont {Jiao}\ \emph {et~al.}(2014)\citenamefont {Jiao},
  \citenamefont {Lau}, \citenamefont {Hatzikirou}, \citenamefont
  {Meyer-Hermann}, \citenamefont {Corbo},\ and\ \citenamefont
  {Torquato}}]{Ji14}%
  \BibitemOpen
  \bibfield  {author} {\bibinfo {author} {\bibfnamefont {Y.}~\bibnamefont
  {Jiao}}, \bibinfo {author} {\bibfnamefont {T.}~\bibnamefont {Lau}}, \bibinfo
  {author} {\bibfnamefont {H.}~\bibnamefont {Hatzikirou}}, \bibinfo {author}
  {\bibfnamefont {M.}~\bibnamefont {Meyer-Hermann}}, \bibinfo {author}
  {\bibfnamefont {J.~C.}\ \bibnamefont {Corbo}}, \ and\ \bibinfo {author}
  {\bibfnamefont {S.}~\bibnamefont {Torquato}},\ }\href@noop {} {\bibfield
  {journal} {\bibinfo  {journal} {Phys. Rev. E}\ }\textbf {\bibinfo {volume}
  {89}},\ \bibinfo {pages} {022721} (\bibinfo {year} {2014})}\BibitemShut
  {NoStop}%
\bibitem [{\citenamefont {Mayer}\ \emph {et~al.}(2015)\citenamefont {Mayer},
  \citenamefont {Balasubramanian}, \citenamefont {Mora},\ and\ \citenamefont
  {Walczak}}]{Ma15}%
  \BibitemOpen
  \bibfield  {author} {\bibinfo {author} {\bibfnamefont {A.}~\bibnamefont
  {Mayer}}, \bibinfo {author} {\bibfnamefont {V.}~\bibnamefont
  {Balasubramanian}}, \bibinfo {author} {\bibfnamefont {T.}~\bibnamefont
  {Mora}}, \ and\ \bibinfo {author} {\bibfnamefont {A.~M.}\ \bibnamefont
  {Walczak}},\ }\href@noop {} {\bibfield  {journal} {\bibinfo  {journal} {Proc.
  Nat. Acad. Sci.}\ }\textbf {\bibinfo {volume} {112}},\ \bibinfo {pages}
  {5950} (\bibinfo {year} {2015})}\BibitemShut {NoStop}%
\bibitem [{\citenamefont {Brauchart}\ \emph {et~al.}(2018)\citenamefont
  {Brauchart}, \citenamefont {Grabner},\ and\ \citenamefont
  {Kusner}}]{Brauchart2018}%
  \BibitemOpen
  \bibfield  {author} {\bibinfo {author} {\bibfnamefont {J.~S.}\ \bibnamefont
  {Brauchart}}, \bibinfo {author} {\bibfnamefont {P.~J.}\ \bibnamefont
  {Grabner}}, \ and\ \bibinfo {author} {\bibfnamefont {W.}~\bibnamefont
  {Kusner}},\ }\href {\doibase 10.1007/s00365-018-9432-8} {\bibfield  {journal}
  {\bibinfo  {journal} {Constr. Approx.}\ } (\bibinfo {year} {2018}),\
  10.1007/s00365-018-9432-8}\BibitemShut {NoStop}%
\bibitem [{\citenamefont {Brauchart}\ \emph {et~al.}()\citenamefont
  {Brauchart}, \citenamefont {Grabner}, \citenamefont {Kusner},\ and\
  \citenamefont {Ziefle}}]{Brauchart2018a}%
  \BibitemOpen
  \bibfield  {author} {\bibinfo {author} {\bibfnamefont {J.~S.}\ \bibnamefont
  {Brauchart}}, \bibinfo {author} {\bibfnamefont {P.~J.}\ \bibnamefont
  {Grabner}}, \bibinfo {author} {\bibfnamefont {W.~B.}\ \bibnamefont {Kusner}},
  \ and\ \bibinfo {author} {\bibfnamefont {J.}~\bibnamefont {Ziefle}},\
  }\href@noop {} {\ }\Eprint {http://arxiv.org/abs/arXiv:1809.02645v1}
  {arXiv:1809.02645v1} \BibitemShut {NoStop}%
\bibitem [{\citenamefont {Brauchart}\ \emph {et~al.}(2014)\citenamefont
  {Brauchart}, \citenamefont {Saff}, \citenamefont {Sloan},\ and\ \citenamefont
  {Womersley}}]{Brauchart2014}%
  \BibitemOpen
  \bibfield  {author} {\bibinfo {author} {\bibfnamefont {J.~S.}\ \bibnamefont
  {Brauchart}}, \bibinfo {author} {\bibfnamefont {E.~B.}\ \bibnamefont {Saff}},
  \bibinfo {author} {\bibfnamefont {I.~H.}\ \bibnamefont {Sloan}}, \ and\
  \bibinfo {author} {\bibfnamefont {R.~S.}\ \bibnamefont {Womersley}},\ }\href
  {\doibase 10.1090/s0025-5718-2014-02839-1} {\bibfield  {journal} {\bibinfo
  {journal} {Mathematics of Computation}\ }\textbf {\bibinfo {volume} {83}},\
  \bibinfo {pages} {2821} (\bibinfo {year} {2014})}\BibitemShut {NoStop}%
\bibitem [{\citenamefont {Marques}\ \emph {et~al.}(2015)\citenamefont
  {Marques}, \citenamefont {Bouville}, \citenamefont {Santos},\ and\
  \citenamefont {Bouatouch}}]{Marques2015}%
  \BibitemOpen
  \bibfield  {author} {\bibinfo {author} {\bibfnamefont {R.}~\bibnamefont
  {Marques}}, \bibinfo {author} {\bibfnamefont {C.}~\bibnamefont {Bouville}},
  \bibinfo {author} {\bibfnamefont {L.~P.}\ \bibnamefont {Santos}}, \ and\
  \bibinfo {author} {\bibfnamefont {K.}~\bibnamefont {Bouatouch}},\ }\href
  {\doibase 10.2200/s00649ed1v01y201505cgr019} {\bibfield  {journal} {\bibinfo
  {journal} {Synthesis Lectures on Computer Graphics and Animation}\ }\textbf
  {\bibinfo {volume} {7}},\ \bibinfo {pages} {1} (\bibinfo {year}
  {2015})}\BibitemShut {NoStop}%
\bibitem [{\citenamefont {Irvine}\ \emph {et~al.}(2012)\citenamefont {Irvine},
  \citenamefont {Bowick},\ and\ \citenamefont {Chaikin}}]{irvine2012}%
  \BibitemOpen
  \bibfield  {author} {\bibinfo {author} {\bibfnamefont {W.~T.}\ \bibnamefont
  {Irvine}}, \bibinfo {author} {\bibfnamefont {M.~J.}\ \bibnamefont {Bowick}},
  \ and\ \bibinfo {author} {\bibfnamefont {P.~M.}\ \bibnamefont {Chaikin}},\
  }\href@noop {} {\bibfield  {journal} {\bibinfo  {journal} {Nat. Mater.}\
  }\textbf {\bibinfo {volume} {11}},\ \bibinfo {pages} {948} (\bibinfo {year}
  {2012})}\BibitemShut {NoStop}%
\bibitem [{\citenamefont {Caillol}\ \emph {et~al.}(1981)\citenamefont
  {Caillol}, \citenamefont {Levesque},\ and\ \citenamefont
  {Weis}}]{Caillol1981}%
  \BibitemOpen
  \bibfield  {author} {\bibinfo {author} {\bibfnamefont {J.~M.}\ \bibnamefont
  {Caillol}}, \bibinfo {author} {\bibfnamefont {D.}~\bibnamefont {Levesque}}, \
  and\ \bibinfo {author} {\bibfnamefont {J.~J.}\ \bibnamefont {Weis}},\ }\href
  {\doibase 10.1080/00268978100102761} {\bibfield  {journal} {\bibinfo
  {journal} {Mol. Phys.}\ }\textbf {\bibinfo {volume} {44}},\ \bibinfo {pages}
  {733} (\bibinfo {year} {1981})}\BibitemShut {NoStop}%
\bibitem [{\citenamefont {Tegmark}(1996)}]{Tegmark1996}%
  \BibitemOpen
  \bibfield  {author} {\bibinfo {author} {\bibfnamefont {M.}~\bibnamefont
  {Tegmark}},\ }\href@noop {} {\bibfield  {journal} {\bibinfo  {journal} {The
  Astrophysical Journal}\ }\textbf {\bibinfo {volume} {470}},\ \bibinfo {pages}
  {L81 } (\bibinfo {year} {1996})}\BibitemShut {NoStop}%
\bibitem [{\citenamefont {Swinbank}\ and\ \citenamefont
  {Purser}(2006)}]{Swinbank2006}%
  \BibitemOpen
  \bibfield  {author} {\bibinfo {author} {\bibfnamefont {R.}~\bibnamefont
  {Swinbank}}\ and\ \bibinfo {author} {\bibfnamefont {R.}~\bibnamefont
  {Purser}},\ }\href {\doibase 10.1256/qj.05.227} {\bibfield  {journal}
  {\bibinfo  {journal} {Q. J. R. Meteorol. Soc.}\ }\textbf {\bibinfo {volume}
  {132}},\ \bibinfo {pages} {1769 } (\bibinfo {year} {2006})}\BibitemShut
  {NoStop}%
\bibitem [{\citenamefont {Weisstein}(2018)}]{SpherePick}%
  \BibitemOpen
  \bibfield  {author} {\bibinfo {author} {\bibfnamefont {E.}~\bibnamefont
  {Weisstein}},\ }\href@noop {} {\enquote {\bibinfo {title} {Sphere point
  picking},}\ }\bibinfo {howpublished} {From MathWorld. A Wolfram Web
  Resource.\url{http://mathworld.wolfram.com/SpherePointPicking.html}}
  (\bibinfo {year} {2018})\BibitemShut {NoStop}%
\bibitem [{\citenamefont {Baddeley}(2004)}]{Baddeley2004}%
  \BibitemOpen
  \bibfield  {author} {\bibinfo {author} {\bibfnamefont {A.}~\bibnamefont
  {Baddeley}},\ }\enquote {\bibinfo {title} {Stochastic geometry},}\ \
  (\bibinfo  {publisher} {Springer},\ \bibinfo {year} {2004})\ Chap.\ \bibinfo
  {chapter} {Spatial Point Processes and their applications}, pp.\ \bibinfo
  {pages} {1 -- 75}\BibitemShut {NoStop}%
\bibitem [{\citenamefont {Lomba}\ \emph {et~al.}(2018)\citenamefont {Lomba},
  \citenamefont {Weis},\ and\ \citenamefont {Torquato}}]{Lomba2018}%
  \BibitemOpen
  \bibfield  {author} {\bibinfo {author} {\bibfnamefont {E.}~\bibnamefont
  {Lomba}}, \bibinfo {author} {\bibfnamefont {J.-J.}\ \bibnamefont {Weis}}, \
  and\ \bibinfo {author} {\bibfnamefont {S.}~\bibnamefont {Torquato}},\ }\href
  {\doibase 10.1103/PhysRevE.97.010102} {\bibfield  {journal} {\bibinfo
  {journal} {Phys. Rev. E}\ }\textbf {\bibinfo {volume} {97}},\ \bibinfo
  {pages} {010102(R)} (\bibinfo {year} {2018})}\BibitemShut {NoStop}%
\bibitem [{\citenamefont {Lomba}\ \emph {et~al.}(2017)\citenamefont {Lomba},
  \citenamefont {Weis},\ and\ \citenamefont {Torquato}}]{Lomba2017}%
  \BibitemOpen
  \bibfield  {author} {\bibinfo {author} {\bibfnamefont {E.}~\bibnamefont
  {Lomba}}, \bibinfo {author} {\bibfnamefont {J.~J.}\ \bibnamefont {Weis}}, \
  and\ \bibinfo {author} {\bibfnamefont {S.}~\bibnamefont {Torquato}},\ }\href
  {\doibase 10.1103/PhysRevE.96.062126} {\bibfield  {journal} {\bibinfo
  {journal} {Phys. Rev. E}\ }\textbf {\bibinfo {volume} {96}},\ \bibinfo
  {pages} {062126} (\bibinfo {year} {2017})}\BibitemShut {NoStop}%
\end{thebibliography}%

\end{document}